\documentclass{amsart}
\pdfoutput=1

\usepackage[T1]{fontenc}

\usepackage{fullpage,graphicx}
\usepackage[french,ruled]{algorithm2e}

\usepackage{pdfpages}

\usepackage{setspace}
\usepackage{dsfont}
\usepackage{pgf,tikz}
\usetikzlibrary{arrows}
\usepackage{lipsum}
\usepackage{fancyhdr}
\usepackage{ifthen}

\usepackage[utf8]{inputenc}
\DeclareMathAlphabet{\mathcalligra}{T1}{calligra}{m}{n}
\usepackage{frcursive}

\usepackage[english]{babel}

\usepackage{hyperref,amsmath,hyperref,graphics,amsfonts,amssymb,verbatim,dsfont,setspace,textcomp,amsthm}

\begin{document}

\author{Emilie Devijver}
\address{Laboratoire de Mathématiques d'Orsay, Faculté des Sciences d'Orsay, Université Paris-Sud, 91405 Orsay, France}
\email{emilie.devijver@math.u-psud.fr}
\author{Yannig Goude}
\address{EDF R\&D, 1 avenue du général De Gaulle, Clamart, France
\\
Laboratoire de Mathématiques d'Orsay, Faculté des Sciences d'Orsay, Université Paris-Sud, 91405 Orsay, France}
\author{Jean-Michel Poggi}
\address{Laboratoire de Mathématiques d'Orsay, Faculté des Sciences d'Orsay, Université Paris-Sud, 91405 Orsay, France, \\
Université Paris Descartes, 143 avenue de Versailles, 75016 PARIS, France }

\title{Clustering electricity consumers using high-dimensional regression mixture models.}% At most 5 thanks

%
% %\date{...}
% %
 \begin{abstract}
    Massive informations about individual (household, small and medium enterprise) consumption
    are now provided with new metering technologies and the smart grid. Two major exploitations
    of these data are load profiling and forecasting at different scales on the grid. Customer
    segmentation based on load classification is a natural approach for these purposes.

    We propose here a new methodology based on mixture of high-dimensional regression models.
    The novelty of our approach is that we focus on uncovering classes or clusters
    corresponding to different regression models. As a consequence, these classes could then be exploited for profiling as well as forecasting in each class or for bottom-up forecasts in a unified view.
    We consider a real dataset of Irish individual consumers of 4,225 meters, each with 48 half-hourly
    meter reads per day over 1 year: from 1st January 2010 up to 31st December 2010, to demonstrate the feasibility of our approach.

 % We consider a multivariate finite mixture of Gaussian regression models for high-dimensional data, where the number of covariates and the size of the response may be much larger than the sample size.
% We provide an $\ell_1$-oracle inequality satisfied by the Lasso estimator according to the Kullback-Leibler loss.
% This result is an  extension of the $\ell_1$-oracle inequality established by Meynet in \cite{Meynet2} in the multivariate case.
% We focus on the Lasso for its $\ell_1$-regularization properties rather than for the variable selection procedure, as it was done in St\"adler in \cite{Stadler}.
 \end{abstract}
%
%
% \subjclass{62H30}
% %
% \keywords{Finite mixture of multivariate regression model, Lasso, $\ell_1$-oracle inequality}
% %
 \maketitle
% % \tableofcontents
%
 
\section{Introduction}
 
 New metering infrastructures as smart meters provide new and potentially massive informations about individual (household, small and medium enterprise) consumption. As an example, in France,
ERDF (Electricite Reseau Distribution de France the French manager of the public electricity distribution network) deployed 250\,000 smart meters, covering a rural and an urban territory and providing half-hourly household energy used each day. ERDF plans to install 35 millions of them over the French territory by the end of 2020 and exploiting such an amount of data is an exciting but challenging task (see \url{http://www.erdf.fr/Linky}).

Many applications coming from individual data analysis can be found in the literature. The first and most popular one is load profiling. Understanding consumers time of use, seasonal patterns and the different features that drive their consumption is a fundamental task for electricity providers to design their offer and more generally for marketing studies (see e.g. \cite{Irwin}). Energy agencies and states can also benefit from profiling for efficiency programs and improve recommendation policies. Customer segmentation based on load classification is a natural approach for that and \cite{leZhou} proposes a nice review of the most popular methods, concluding that classification for smart grids is a hard task due to the complexity, massiveness, high dimension and heterogeneity of the data. Another problem pointed out is the dynamic structure of smart meters data and particularly the issue of portfolio variations (losses and gains of customers), the update of a previous classification when new customers arrive or the clustering of a new customer with very few observations on its load. In \cite{Kwac}, the authors propose a segmentation algorithm based on K-means to uncover shape dictionaries that help to summarize information and cluster a large population of 250\,000 households in California. However, the proposed solution exploits a quite long historic of data of at least 1 year. 

Recently, other important questions were raised by smart meters and the new possibility to send potentially complex signal to consumers (incentive payments, time varying prices...) and demand response program tailoring attracts a lot of attention (see \cite{US}, \cite{Hancher}). Local optimization of electricity production and real time management of individual demand thus play an important role in the smart grid landscape. It induces a need for local electricity load forecasting at different levels of the grid and favorites bottom-up approaches based on a two stage process. First, it consists in building classes in a population such that each class could be sufficiently well forecast but corresponds to different load shapes or reacts differently to exogenous variables like temperature or prices (see e.g. \cite{Labeeuw} in the context of demand response). The second stage consists in aggregating forecasts to forecast the total or any subtotal of the population consumption. For example, identify and forecast the consumption of a sub-population reactive to an incentive is an important need to optimize a demand response program. Surprisingly, few papers consider the problem of clustering individual consumption for forecasting and specially for forecasting at a disaggregated level (e. g. in each class). In \cite{Alzate}, clustering procedures are compared according to the forecasting performances of their corresponding bottom-up forecasts of the total consumption of 6\,000 residential customers and small-to-medium enterprises in Ireland. Even if they achieve nice performances at the end, the proposed clustering methods are quite independent to the VAR model used for forecasting. In \cite{MisitiElec}, a clustering algorithm is proposed that couples hierarchical clustering and multi-linear regression models to improve the forecast of the total consumption of a French industrial subset. They obtain a real forecasting gain but need a sufficiently long dataset (2-3 years) and the algorithm is computationally intensive.

We propose here a new methodology based on high dimensional regression models. Our main contribution is that we focus on uncovering classes corresponding to different regression models. As a consequence, these classes could then be exploited for profiling as well as forecasting in each classes or for bottom-up forecasts in an unified view. More precisely, we consider regression models where $Y_d=X_d\beta+\varepsilon_d$ is typically an individual -high dimension- load curve for day $d$ and $X_d$ could be alternatively $Y_{d-1}$ or any other exogenous covariates.  

We consider a real dataset of Irish individual consumers of 4\,225 meters, each with 48 half-hourly meter reads per day over 1 year: from 1st January 2010 up to 31st December 2010. These data have already been studied in \cite{Alzate} and \cite{Chaouch2014} and we refer to those papers for a presentation of the data. For computational and time reasons, we draw a random sample of around 500 residential consumers among the $90 \%$ closest to the mean, to demonstrate the feasibility of our methods. We show that, considering only 2 days of consumption, we obtain physically interpretable clusters of consumers.

\begin{figure}[!ht]
\centering
   \includegraphics[scale=0.18]{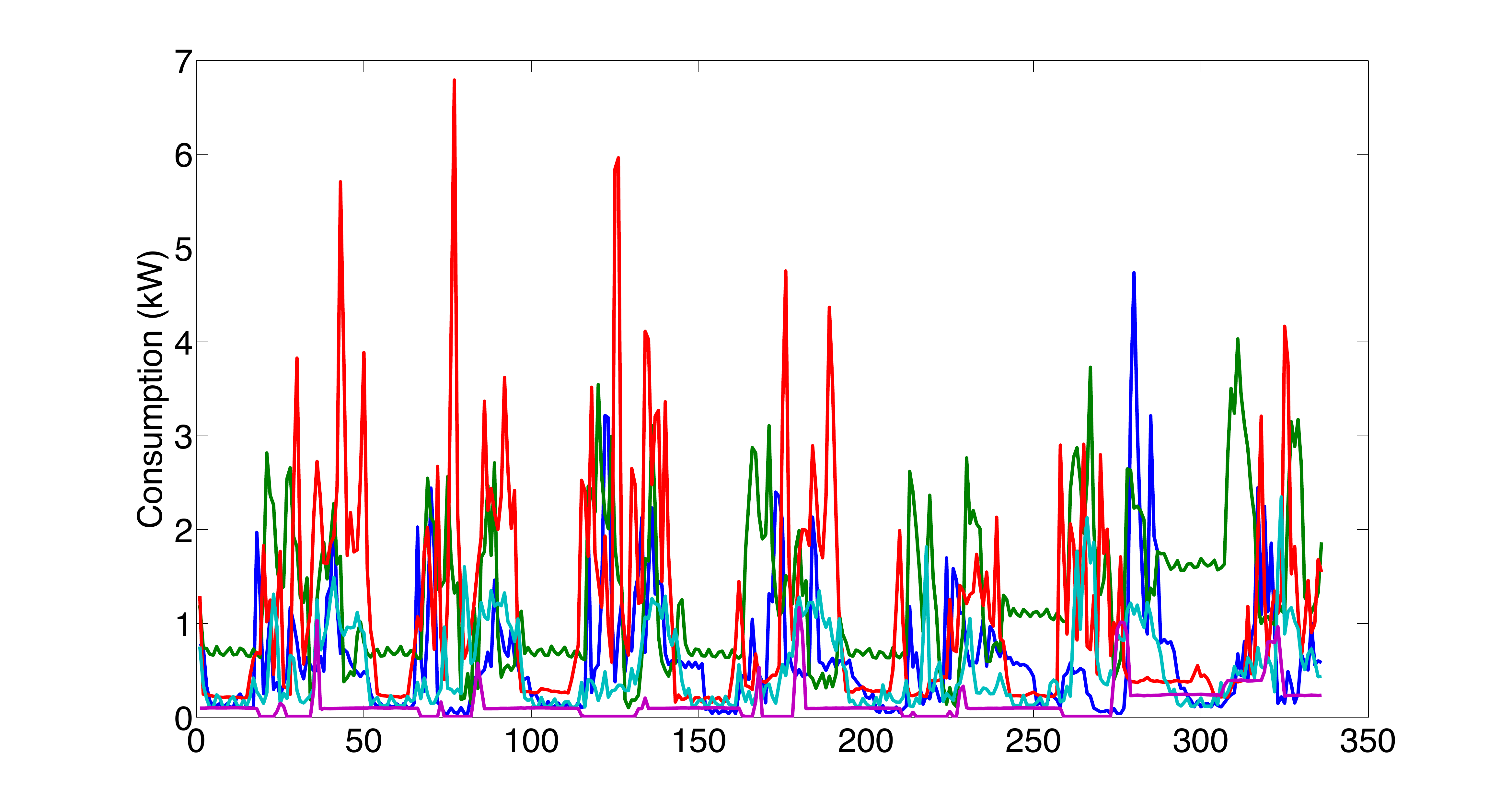}
   \caption{Load consumption of a sample of $5$ consumers over a week in winter}
   \label{sample}
\end{figure}

According to the Fig. \ref{sample}, deal with individual consumption curves is an hard task, because of the high variability.

 \section{Method}
 \label{Method}
 
 We propose to use model-based clustering and adopt the model selection paradigm.
 Indeed, we consider the model collection of conditional mixture densities,
 $$\mathcal{S} = \left\{ s_{\xi}^{(K,J)}, K \in \mathcal{K}, J \in \mathcal{J} \right\},$$
 where $K$ denotes the number of clusters, $J$ the set of relevant variables for clustering, and $\mathcal{K}$ and $\mathcal{J}$ being respectively 
 the set of possible values of
 $K$ and $J$.
 
 The basic model we propose to use is a finite mixture regression of $K$ multivariate Gaussian densities (see \cite{Stadler} for a recent and fruitful reference), the conditional density being, for $x \in \mathbb{R}^p, y \in \mathbb{R}^q$,
 $\varphi$ denoting the Gaussian density,
 \begin{align*}
 s_{\xi}^{(K,J)}(y|x)=&\sum_{k=1}^{K} \pi_k \varphi(\beta_k^{[J]} x, \Sigma_k)
 %\frac{\pi_{k}}{(2 \pi)^{q/2}\text{det}(\Sigma_k)^{1/2}}  \times  e^ {-\frac{(y-\beta^{[J]}_{k} x)^t \Sigma_{k}^{-1}(y-\beta^{[J]}_{k} x)}{2} }.\\
\end{align*}

Such a model can be interpreted and used from two different viewpoints.

\noindent
First, from a clustering perspective, given the estimation $\hat{\xi}$ of the parameters $\xi =( \boldsymbol{\pi},\boldsymbol{\beta},\boldsymbol{\Sigma})$, 
we could deduce data clustering from the Maximum A Posteriori principle: for 
 each observation $i$,
we compute the posterior probability $\tau_{i,k}(\hat{\xi})$  of each cluster $k$ from the estimation $\hat{\xi}$, 
 and we assign the observation $i$ to the cluster $\hat{k}_i = \underset{k \in \{1,\ldots,K\}}{\operatorname{argmax}} \tau_{i,k}(\hat{\xi})$.
 Proportions of each cluster are estimated by $\hat{\pi}$.
 
 \noindent
 Second, in each cluster, the corresponding model is meaningful and its interpretation allows to understand the relationship between variables $Y$ and $X$ 
 since it is of the form
 $$Y=X \beta_k +\epsilon,$$
 the noise intensity being measured by $\Sigma_k$.
 
Parameters are estimated from the Lasso-MLE procedure, which is described in details in \cite{procedures}, and theoretically approved in \cite{inegOracleLassoMLE}.
 To overcome the high-dimension issue, we use the Lasso estimator on the regression parameters and we restrict the covariance matrix to be diagonal.
To avoid shrinkage, we estimate parameters by Maximum Likelihood Estimator on relevant variables selected by the Lasso estimator.
Rather than selecting a regularization parameter, we present this issue at a model selection problem, considering a grid of regularization parameters.
Indices of relevant variables for this grid of regularization parameters are denoted by $\mathcal{J}$.
Since we also have to estimate the number of components, we compute those models for different number of components, belonging to $\mathcal{K}$.
In this paper, $\mathcal{K} = \{1,\ldots,8\}$.

 Among this collection, we could focus on a few models which seem interesting for clustering, depending on which characteristics we want to highlight.
 We propose to use the slope heuristics to extract potentially interesting models.
 The selected  model  minimizes the log-likelihood penalized by $2 \hat{\kappa} D_{(K,J)}/n$, where $D_m$ denotes the dimension of the model $m$, and
 where $\hat{\kappa}$ is constructed from a completely  data-driven procedure. In practice, we use the Capushe package, see \cite{baudry}.
 
 In addition to this family of models, we need to have powerful tools to translate curves into variables.
 Rather than dealing with the discretization of the load consumption, we project it onto a functional basis to take into
 account the functional structure.
 Since we are interested in not only representing the curves into a functional basis, but also to benefit from a time-scale interpretation of coefficients, 
 we propose to use wavelet basis, see \cite{Mallat} for a theoretical approach, and  \cite{Misiti} for a practical purpose.
 To simplify our presentation, we will focus on the Haar basis.

 \section{Typical workflow using the example of the aggregated consumption}
 \subsection{General framework}
\begin{figure}[!ht]
 \begin{center}
 \begin{minipage}[c]{.25\linewidth}
 \centering
   \includegraphics[scale=0.21]{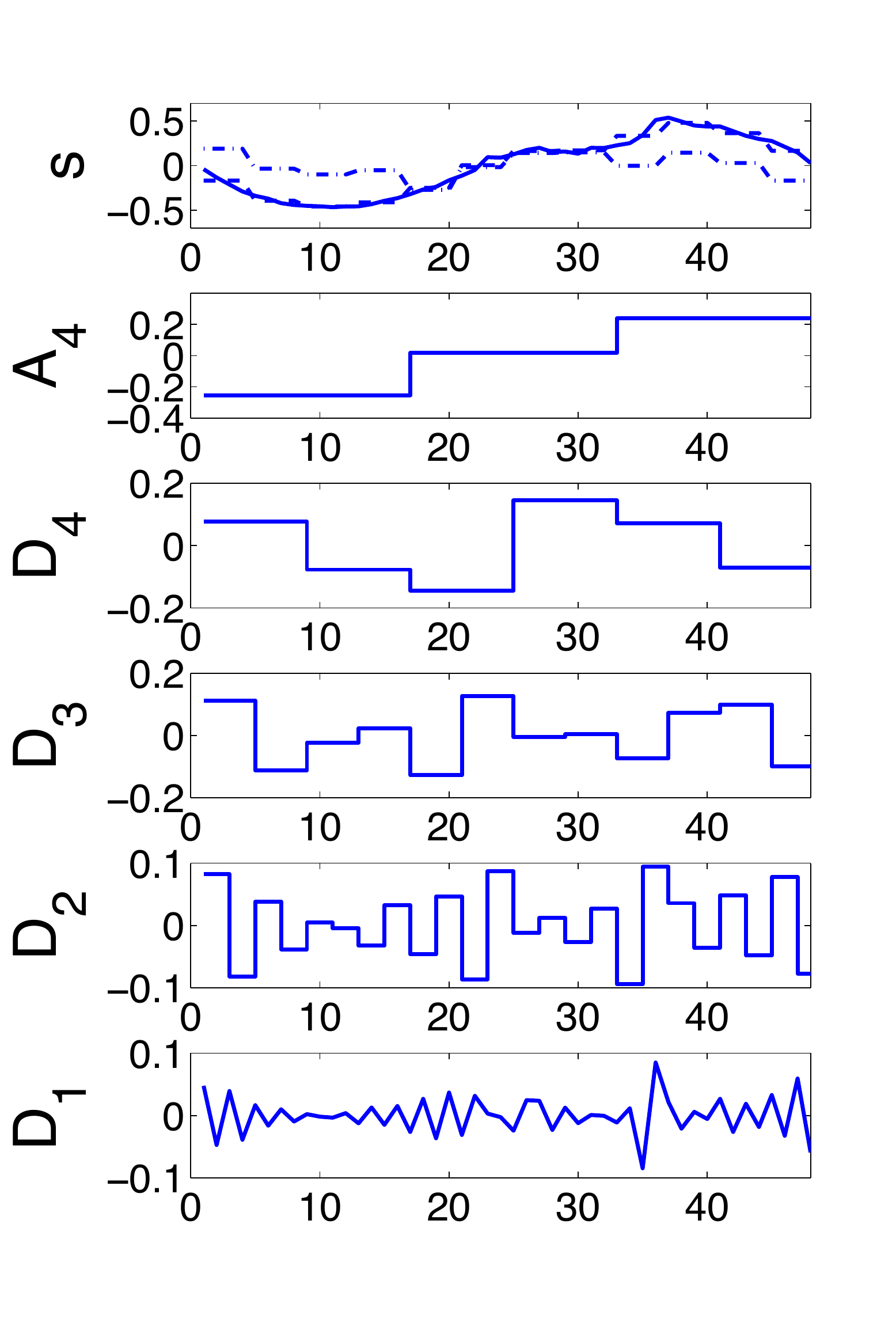}
 \end{minipage}
 %\hfill
 \begin{minipage}[c]{.4\linewidth}
\centering
 \includegraphics[scale=0.2]{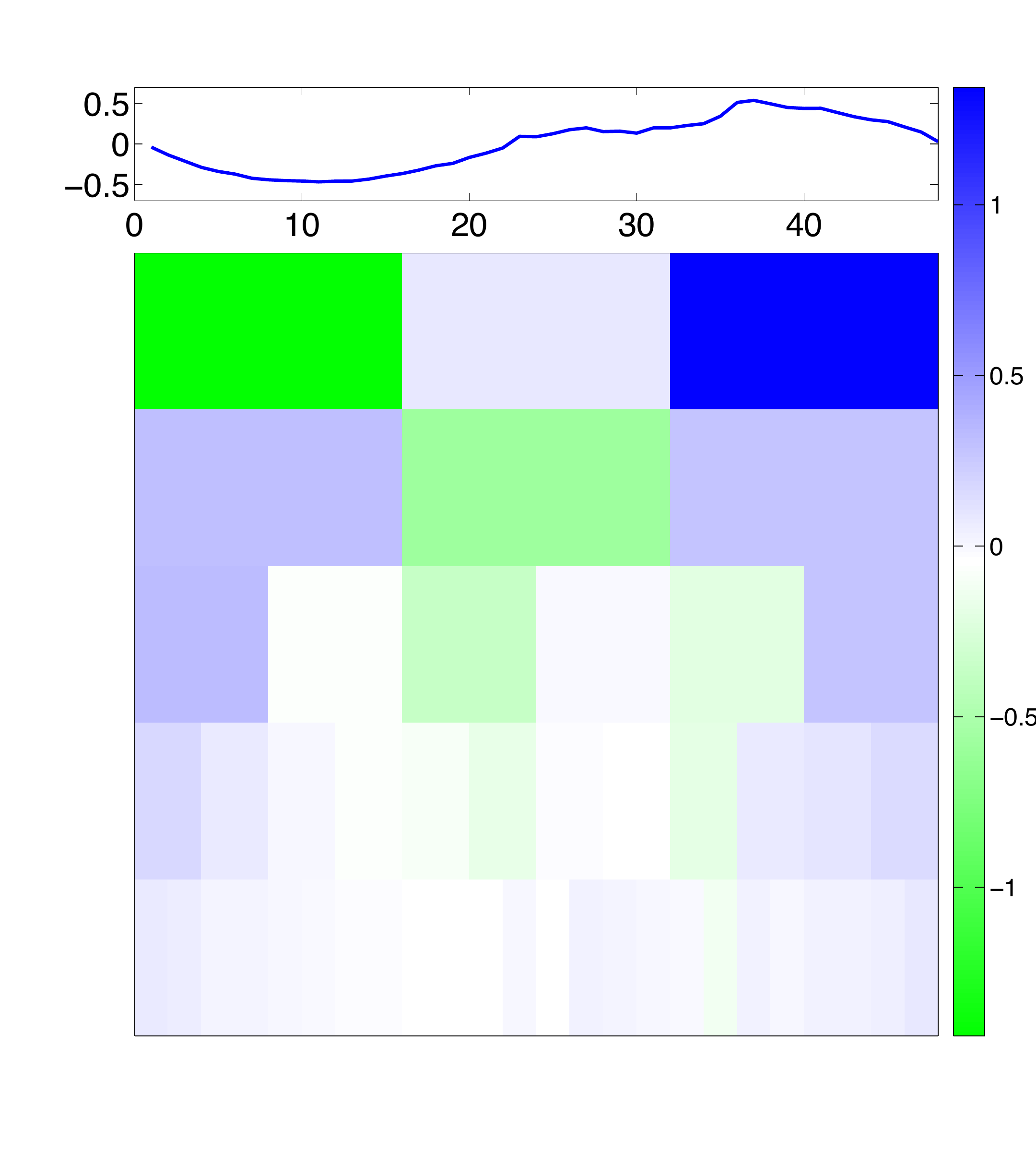}
 \end{minipage}
 \caption{Projection of a load consumption for one day into Haar basis, level $4$. By construction, we get $s=A_4+D_4+D_3+D_2+D_1$.
 On the left side, the signal is considered with reconstruction of dataset, the dotted being preprocessing $1$ and the dotted-dashed being the preprocessing $2$}
 \label{decompositionOndelettes}
  
 \end{center}

 \end{figure}

 The goal is to cluster electricity consumers using a regression mixture model.
 We will consider the consumption of the eve for the regressors, to explain the consumption of the day.
 Consider the daily consumption, where we observe $48$ points.
 We project the signal onto the Haar basis at level $4$.
 The signal could be decomposed in approximation, denoted by $A_4$, and several details, denoted by $D_4$, $D_3$, $D_2$, and $D_1$.
 We illustrate it in Fig. \ref{decompositionOndelettes}, where in addition the decomposition in  sum of orthogonal signals on the left,
 one can find a colored representation of the corresponding wavelet coefficients in the time scale plane.
For an automatic denoising, we remove details of level $1$ and $2$, which correspond to high-frequency components.
Two centerings will be considered:
 \begin{itemize}
  \item preprocessing 1: before projecting, we center each signal individually. 
  \item preprocessing 2: we consider details coefficients of level $4$ and $3$. Here, we remove also a low-frequency approximation.
 \end{itemize}
 Depending on the preprocessing, we will get different clusterings

 We observe the load consumption of $n$ residentials over a year, denoted by $(z_{i,t})_{1\leq i \leq n, t\in T}$.
 We consider 
 \begin{itemize}
  \item $Z_t = \sum_{i=1}^n z_{i,t}$ the aggregated signal,
  \item $\mathfrak{Z}_{d} = (Z_t)_{48(d-1) \leq t \leq 48d}$ the aggregated signal for the day $d$,
  \item $\mathfrak{z}_{i,d} = (z_{i,t})_{48(d-1) \leq t \leq 48d}$ the signal for the residential $i$ for the day $d$.
 \end{itemize}
 
 We consider three different ways to analyze this dataset.
 
 The first one consider $(\mathfrak{Z}_{d},\mathfrak{Z}_{d+1})_{1 \leq d \leq 338}$ over time, and the results are easy to interpret.
 We take this opportunity to develop in details the steps of the method we propose from the model to the clusters via model visualization
 and interpretation.
 In the second one, we want to cluster consumers on mean days. Working with mean days leads to some stability.
 The last one is the most difficult, since we consider individuals curves $(\mathfrak{z}_{i,d_0},\mathfrak{z}_{i,d_0+1})_{1\leq i \leq n}$ 
 and we classify these individuals for the days $(d_0,d_0+1)$.

 \subsection{Cluster days on the residential synchronous curve}
 \label{ClusterDays}
 In this Section, we focus on the residential synchronous $(Z_t)_{t \in T}$.
 We will illustrate our procedure step by step, and highlight some features of data.
 The whole analysis will be done for the preprocessing $2$.
 \subsubsection{Model selection}
 Our procedure leads to a model collection, with various number of components and various sparsities.
 Let us explain how to select some interesting models, thanks to the slope heuristic.
 We define 
 $$(K(\kappa),J(\kappa)) = \underset{(K,J)}{\operatorname{argmin}} \left(-\gamma_n(\hat{s}^{(K,J)}) + 2\kappa D_{(K,J)}/n\right),$$
 where $\gamma_n$ is the log-likelihood function and $\hat{s}^{(K,J)}$ is the log-likelihood minimizer among the collection $S^{(K,J)}$.
 We consider the step function $\kappa \mapsto D_{K(\kappa),J(\kappa)}$,  $\hat{\kappa}$ being the abscissa which leads to the biggest dimension jump.
 We select the model $(\hat{K},\hat{J}) = (K(2\hat{\kappa}),J(2\hat{\kappa}))$. 
 To improve that, we could consider the points $(D_{(K,J)},-\gamma_n(s^{(K,J)}) + 2\hat{\kappa} D_{(K,J)}/n)_{(K,J)}$, and select some models
 minimizing this criterion.
 \begin{figure}[!ht]
 \begin{minipage}[c]{.32\linewidth}
 \centering
 \includegraphics[scale=0.15]{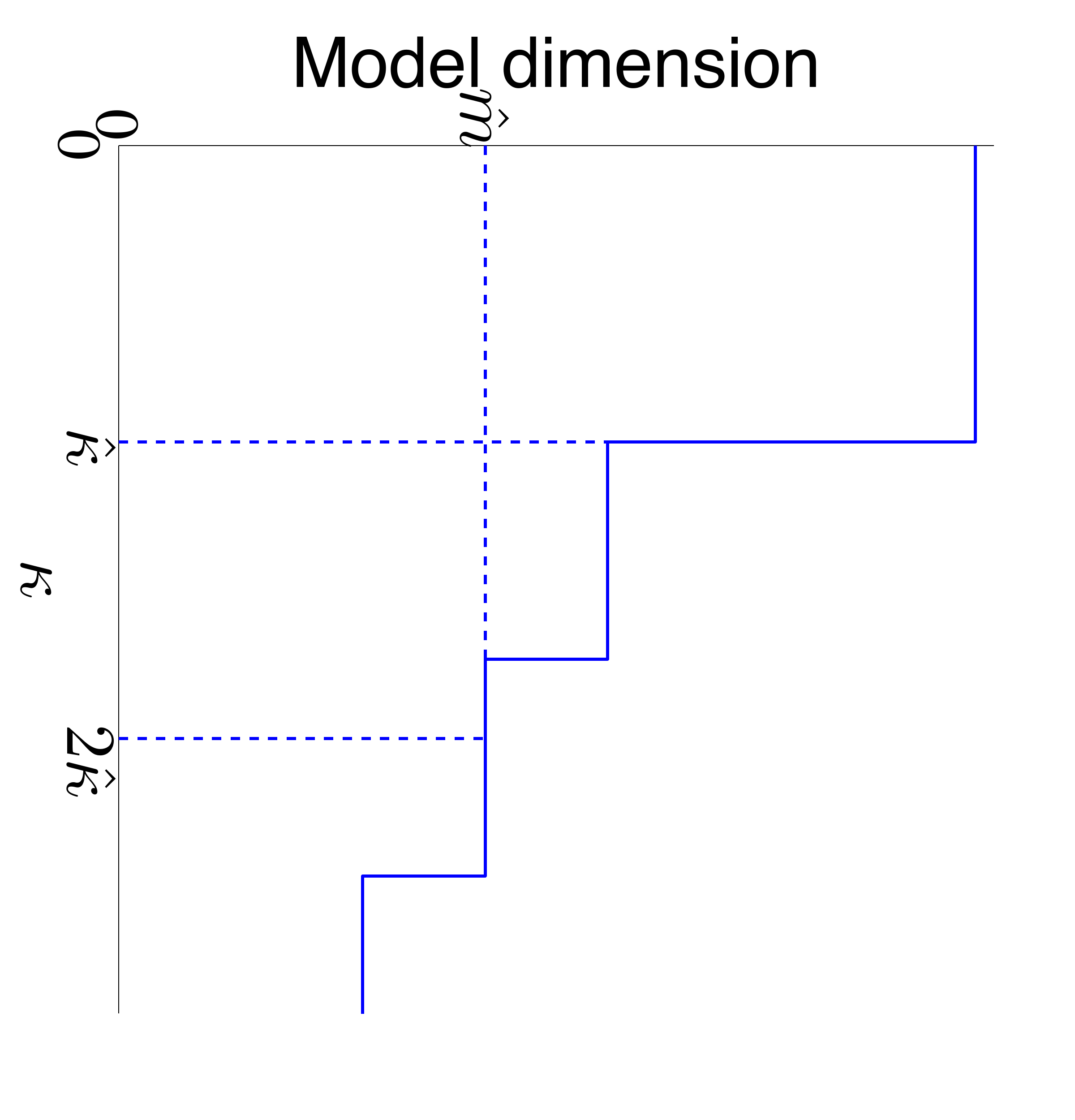}
 \caption{We select the model $\hat{m}$ using the slope heuristic}
 \label{sautDeDimension}
\end{minipage}
\hfill
  \begin{minipage}[c]{.65\linewidth}
  \centering
 \includegraphics[scale=0.17]{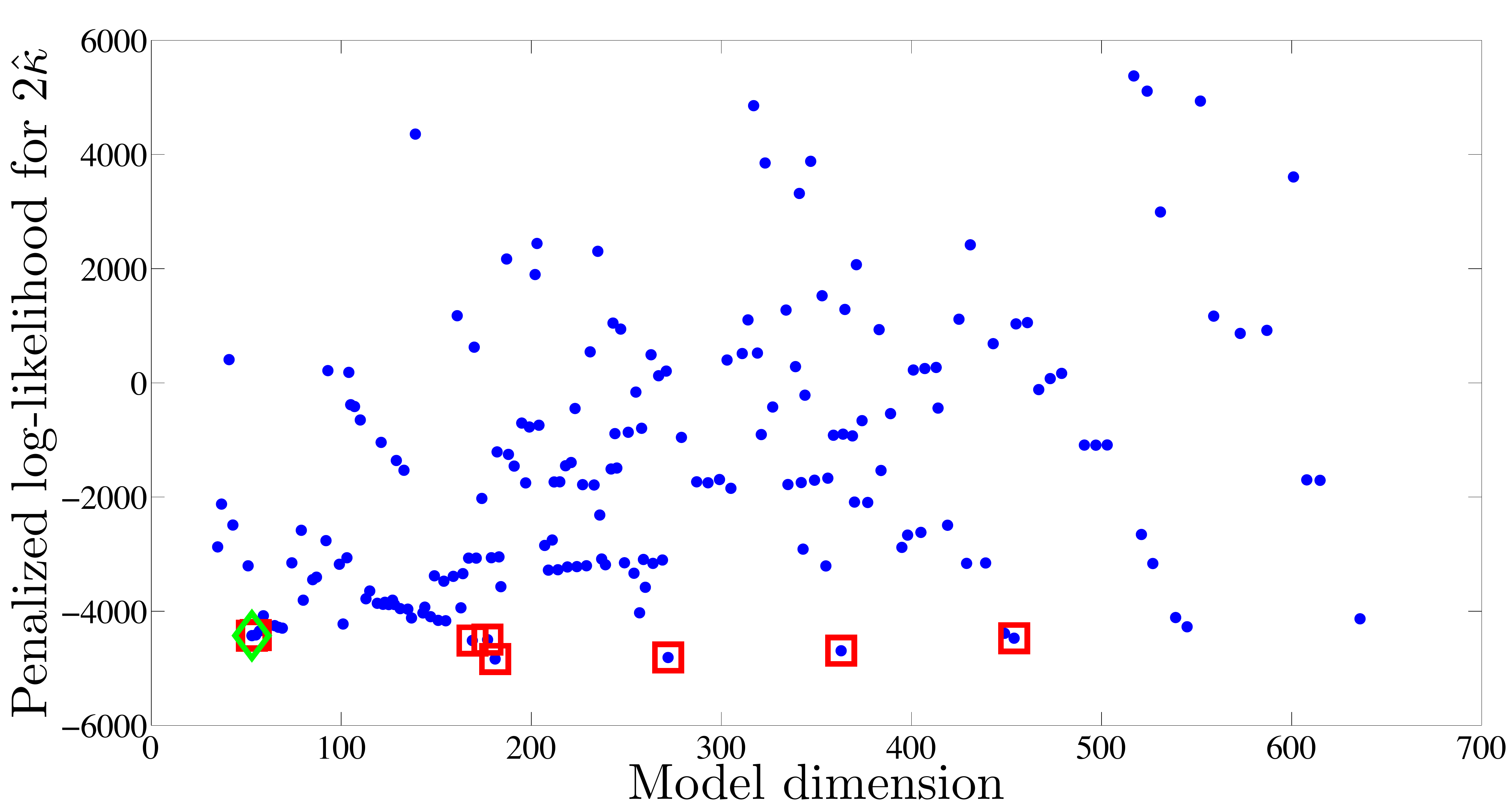}
 \caption{Minimization of the penalized log-likelihood. Interesting models are branded by red squares, the selected one by green diamond}
\label{pentes}
\end{minipage}
\end{figure}
According to Figs. \ref{sautDeDimension} and \ref{pentes}, we could consider some $\hat{\kappa}$ which seem to create big jumps, and several models which seem to minimize the penalized log-likelihood.

 \subsubsection{Model visualization}
\begin{figure}[!ht]
 \begin{minipage}[c]{.32\linewidth}
 \centering
  \includegraphics[scale=0.18]{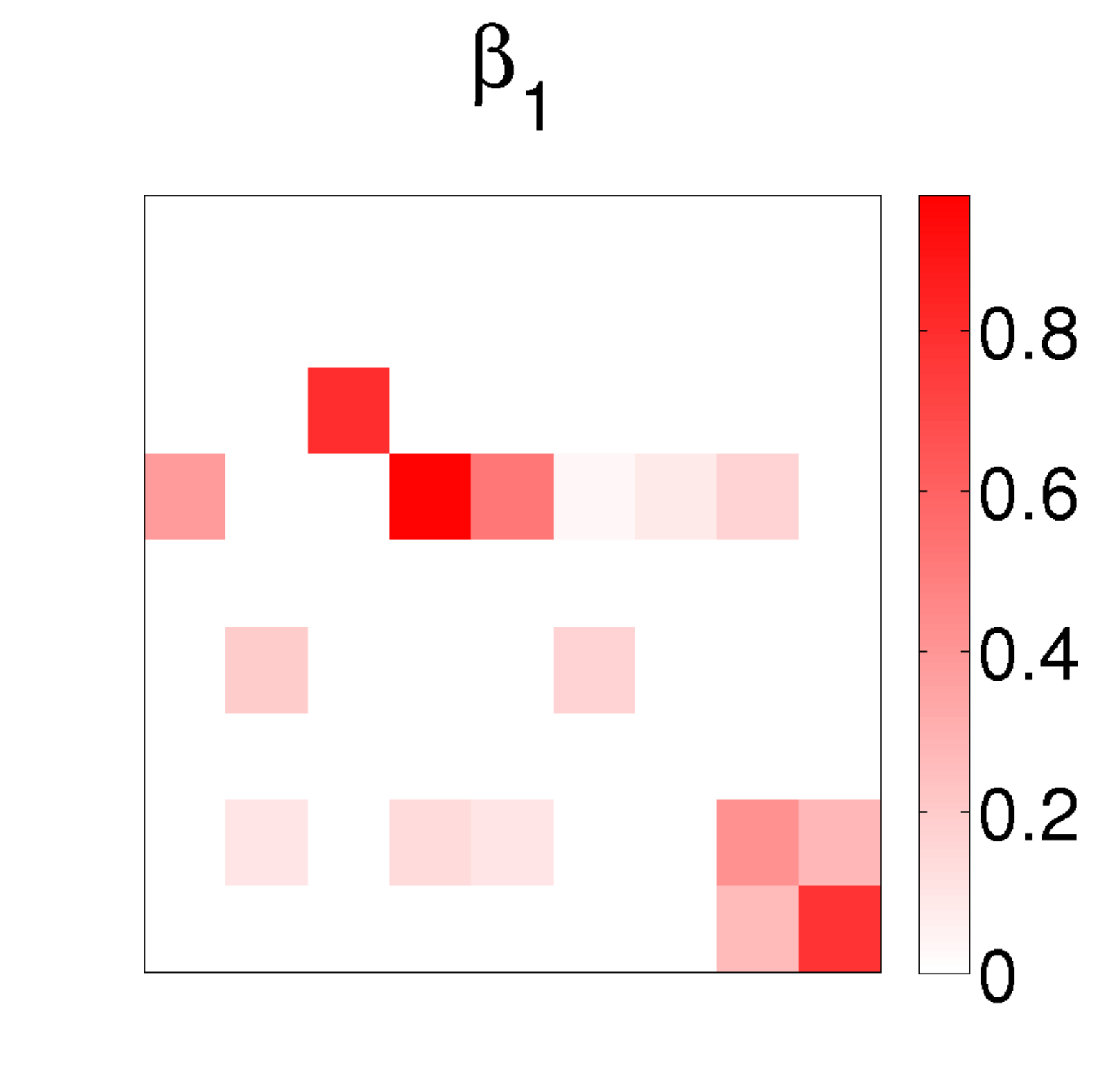}
 \end{minipage}
 \hfill
  \begin{minipage}[c]{.32\linewidth}
 \centering
 \includegraphics[scale=0.18]{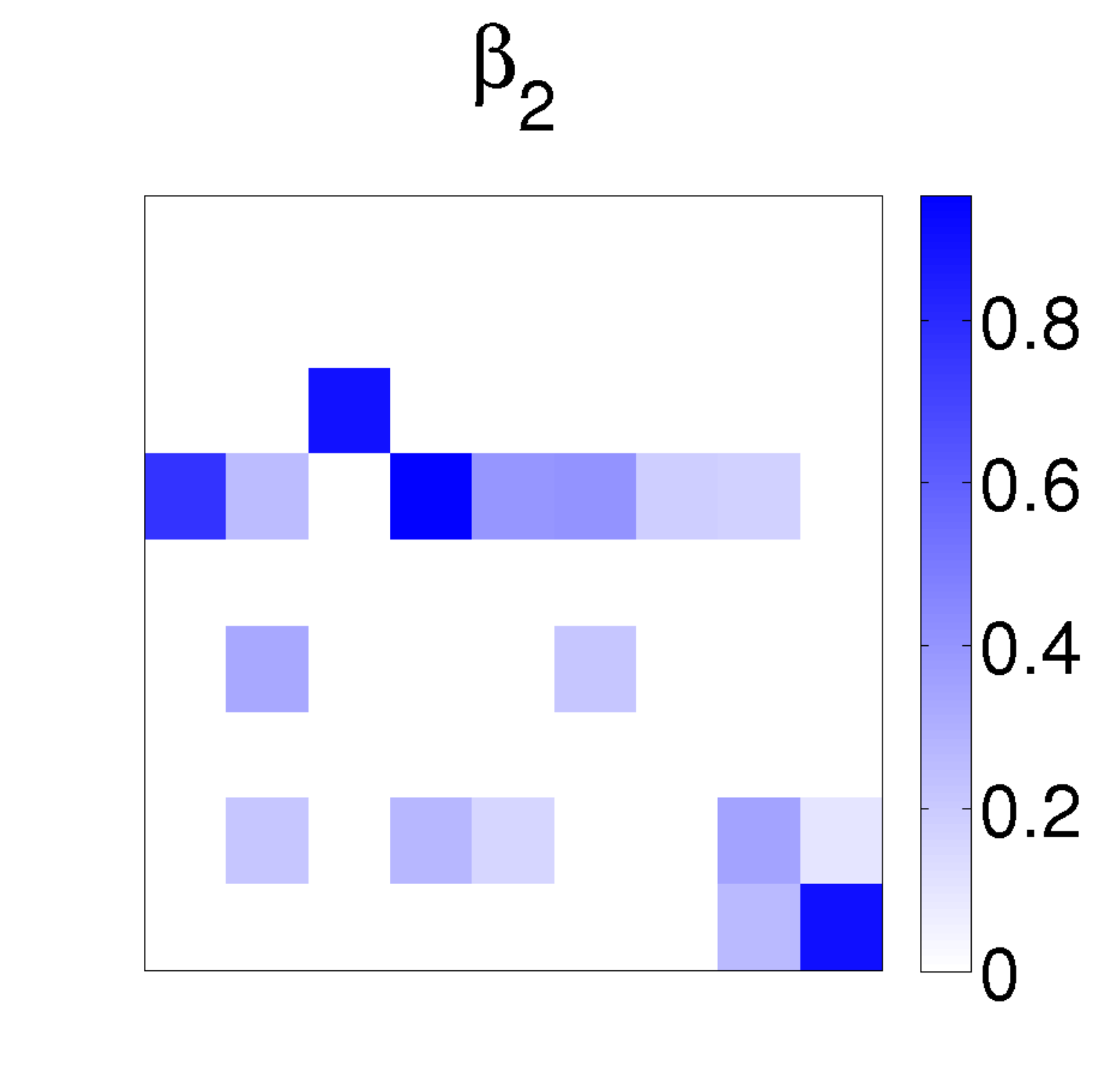}
 \end{minipage}
  \hfill
  \begin{minipage}[c]{.32\linewidth}
  \centering
  \includegraphics[scale=0.18]{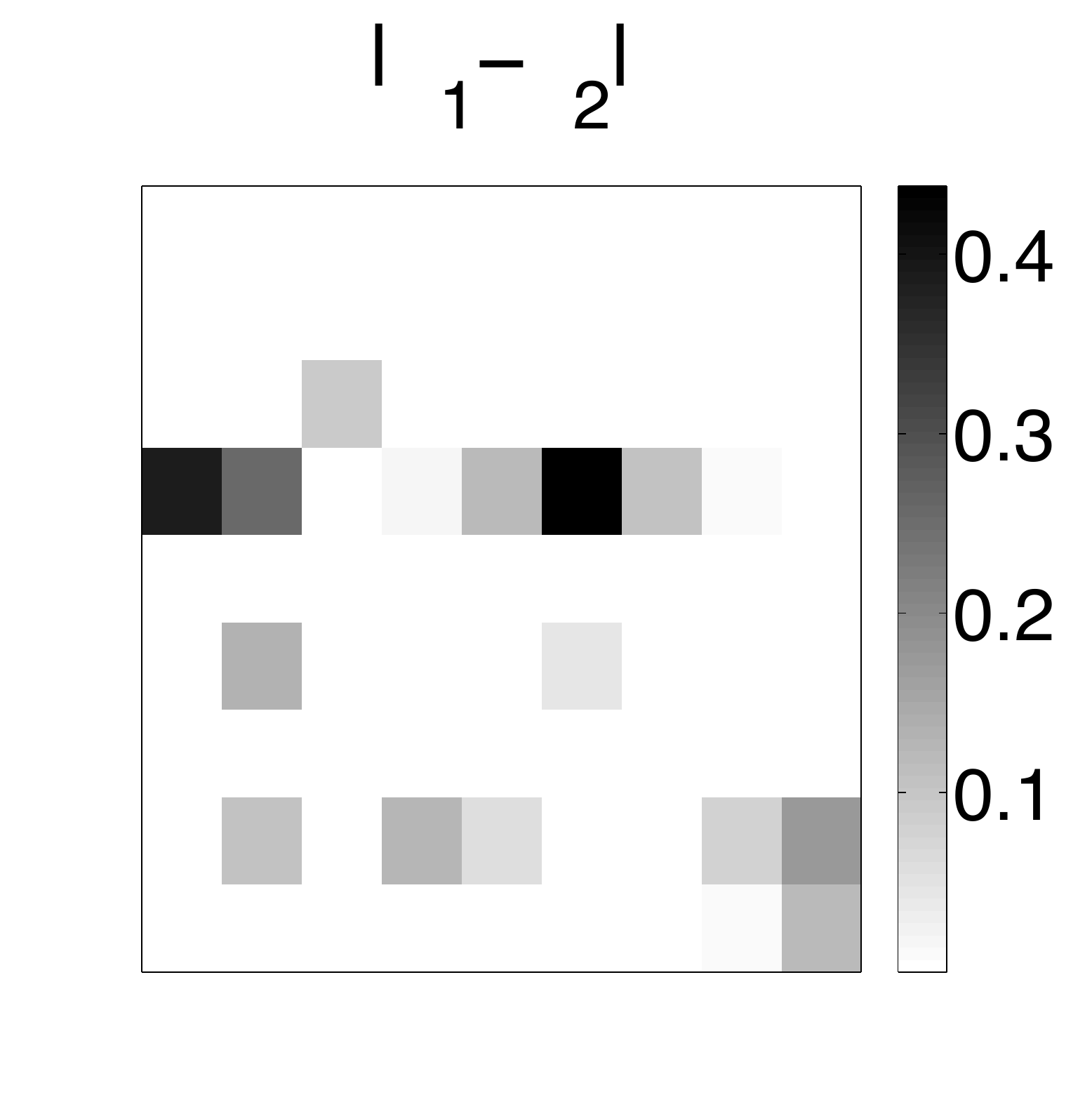}
 \end{minipage}
 \caption[For the selected model, we represent $\hat{\beta}$ in each cluster]{For the selected model, we represent $\hat{\beta}$ in each cluster.
 Absolute values of coefficients are represented by different colormaps, white for $0$.
 Each color represents a cluster
 }
 \label{beta}
\end{figure}

 Thanks to the model-based clustering, we have constructed a model in each cluster.
 Then, we could understand differences between clusters from $\hat{\beta}$ and $\hat{\Sigma}$ estimations.
 We represent it with an image, each coefficient being represented by a pixel.
 As we consider the linear model $Y = X \beta_k $ for each cluster, rows correspond to regressors coefficients and  columns to response coefficients.
Diagonal coefficients will explain the main part.
 The Figs. \ref{dessinTikz} and \ref{dessinTikz2} explain the image construction, whereas we compute it for the model selected 
 by the previous step in Fig. \ref{beta}.

\begin{figure}[!ht]
\begin{minipage}[c]{.48 \linewidth}
\centering
 \definecolor{rouge}{rgb}{1,0.2,0.2}
\definecolor{gris}{rgb}{0.2,0.2,0.2}
\begin{tikzpicture}[scale=0.6]
\clip(-2.6,-2.6) rectangle (3.5,2.2);
\fill[color=gris,fill=gris,fill opacity=0.1] (-1,2) -- (-1,-2) -- (3,-2) -- (3,2) -- cycle;
\fill[color=rouge,fill=rouge,fill opacity=0.1] (0,2) -- (-1,2) -- (-1,1) -- (0,1) -- cycle;
\fill[color=rouge,fill=rouge,fill opacity=0.1] (0,1) -- (0,0) -- (1,0) -- (1,1) -- cycle;
\fill[color=rouge,fill=rouge,fill opacity=0.1] (1,0) -- (1,-2) -- (3,-2) -- (3,0) -- cycle;
\draw [color=gris] (-1,2)-- (-1,-2);
\draw [color=gris] (-1,-2)-- (3,-2);
\draw [color=gris] (3,-2)-- (3,2);
\draw [color=gris] (3,2)-- (-1,2);
\draw (1,2)-- (1,-2);
\draw (0,2)-- (0,-2);
\draw (-1,0)-- (3,0);
\draw (-1,1)-- (3,1);
\draw (-2,1.8) node[anchor=north west] {$A_4$};
\draw (-2,0.8) node[anchor=north west] {$D_4$};
\draw (-2,-0.54) node[anchor=north west] {$D_3$};
\draw (-01.1,-1.85) node[anchor=north west] {$A_4$};
\draw (0,-1.85) node[anchor=north west] {$D_4$};
\draw (1.5,-1.85) node[anchor=north west] {$D_3$};
\draw [color=rouge] (0,2)-- (-1,2);
\draw [color=rouge] (-1,2)-- (-1,1);
\draw [color=rouge] (-1,1)-- (0,1);
\draw [color=rouge] (0,1)-- (0,2);
\draw [color=rouge] (0,1)-- (0,0);
\draw [color=rouge] (0,0)-- (1,0);
\draw [color=rouge] (1,0)-- (1,1);
\draw [color=rouge] (1,1)-- (0,1);
\draw [color=rouge] (1,0)-- (1,-2);
\draw [color=rouge] (1,-2)-- (3,-2);
\draw [color=rouge] (3,-2)-- (3,0);
\draw [color=rouge] (3,0)-- (1,0);
\draw [->,line width=1pt] (-1.5,-1.5) to[bend right] (-1,-2.4);
\end{tikzpicture}

\caption{Representation of the regression matrix $\beta_k$ for the preprocessing $1$. 
}
\label{dessinTikz}
\end{minipage}
\hfill
\begin{minipage}[c]{.48 \linewidth}
\centering
 \definecolor{rouge}{rgb}{1,0.2,0.2}
\definecolor{gris}{rgb}{0.2,0.2,0.2}
\begin{tikzpicture}[scale=0.6]
\clip(-2.6,-2.6) rectangle (3.5,2.2);
\fill[color=gris,fill=gris,fill opacity=0.1] (-1,2) -- (-1,-2) -- (3,-2) -- (3,2) -- cycle;
%\fill[color=rouge,fill=rouge,fill opacity=0.1] (0,2) -- (-1,2) -- (-1,1) -- (0,1) -- cycle;
\fill[color=rouge,fill=rouge,fill opacity=0.1] (-1,2) -- (1/3,2) -- (1/3,2/3) -- (-1,2/3) -- cycle;
\fill[color=rouge,fill=rouge,fill opacity=0.1] (1/3,2/3) -- (3,2/3) -- (3,-2) -- (1/3,-2) -- cycle;
\draw [color=gris] (-1,2)-- (-1,-2);
\draw [color=gris] (-1,-2)-- (3,-2);
\draw [color=gris] (3,-2)-- (3,2);
\draw [color=gris] (3,2)-- (-1,2);
\draw (1/3,2)-- (1/3,-2);
\draw (-1,2/3)-- (3,2/3);
\draw (-2,1.6) node[anchor=north west] {$D_4$};
\draw (-2,-0.5) node[anchor=north west] {$D_3$};
\draw (-1,-1.85) node[anchor=north west] {$D_4$};
\draw (1.2,-1.85) node[anchor=north west] {$D_3$};
\draw [color=rouge] (1/3,2)-- (-1,2);
\draw [color=rouge] (-1,2)-- (-1,2/3);
\draw [color=rouge] (-1,2/3)-- (1/3,2/3);
\draw [color=rouge] (1/3,2)-- (1/3,-2);
\draw [color=rouge] (1/3,-2)-- (3,-2);
\draw [color=rouge] (3,-2)-- (3,0);
\draw [color=rouge] (3,2/3)-- (1/3,2/3);
\draw [->,line width=1pt] (-1.5,-1.5) to[bend right] (-1,-2.4);
\end{tikzpicture}

\caption{Representation of the regression matrix $\beta_k$ for the preprocessing $2$.}
\label{dessinTikz2}
\end{minipage}

\end{figure}

To highlight differences between clusters, we also plot $\hat{\beta}_1-\hat{\beta}_2$.
First, we remark that $\hat{\beta}_1$ and $\hat{\beta}_2$ are sparse, thanks to the Lasso estimator.
Moreover, the main difference between $\hat{\beta}_1$ and $\hat{\beta}_2$ is row $4$, columns $1, 2$ and $6$.
We could say that the procedure uses, depending on cluster, more or less the first coefficient of $D_3$ of $X$ to describe
coefficients $1$ and $2$ of $D_3$ and coefficient $3$ of $D_4$ of $Y$.
The Fig. \ref{courbesMoyennesPT2}  enlightens those differences between clusters.

\begin{figure}[!ht]
\centering
 \includegraphics[scale=0.12]{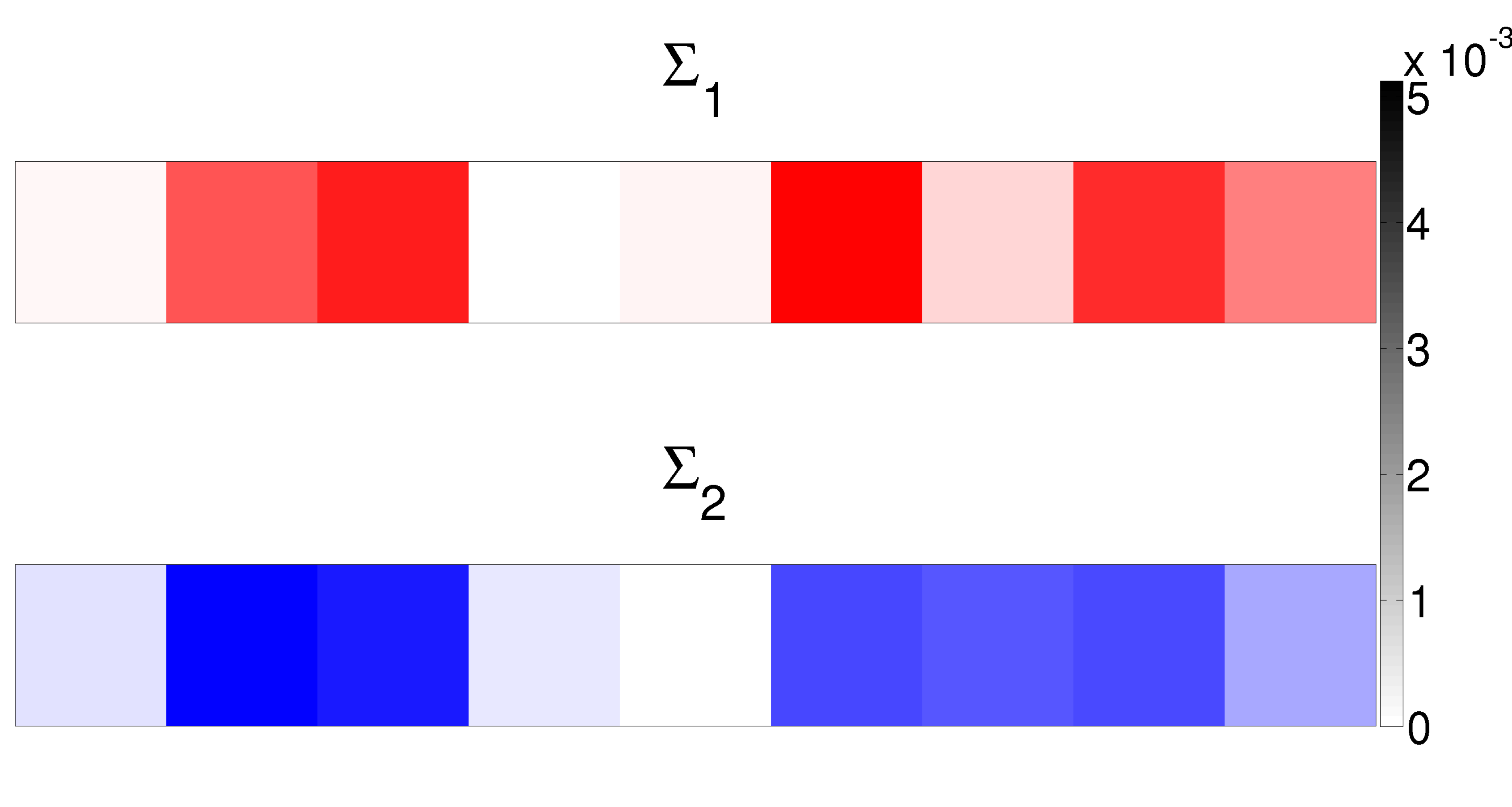}
 \caption{For the model selected, we represent $\Sigma$ in each cluster}
\label{sigma}
\end{figure}

We represent the covariance matrix in Fig. \ref{sigma}. Because we estimate it by a diagonal matrix in each cluster, we just display the diagonal coefficients.
We keep the same scale for all the clusters, to highlight which clusters are noisier.

 \subsubsection{Model-based clustering}
\begin{figure}[!ht]
\centering
 \includegraphics[scale=0.3]{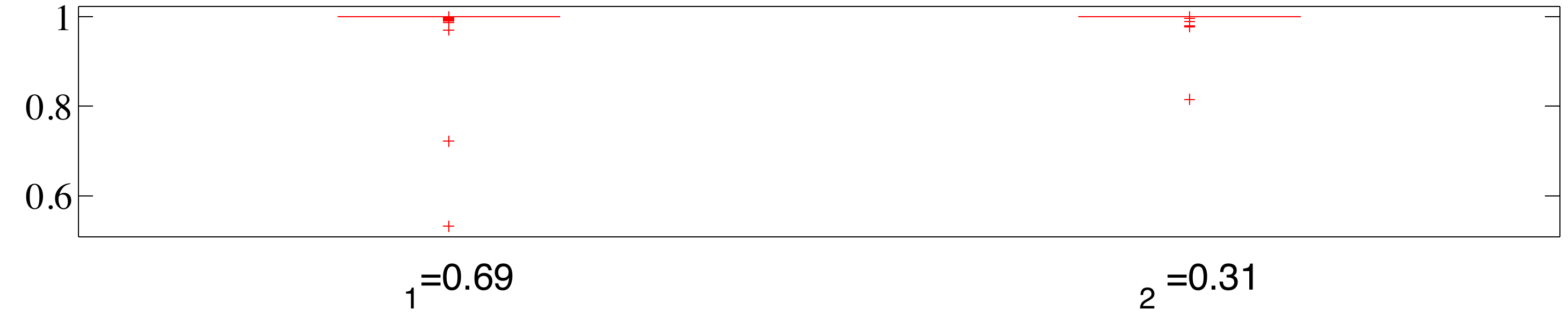}
 \caption{Assignment boxplots per cluster}
 \label{affectation}
\end{figure}

 We could compute the posterior probability for each observation.
 In Fig. \ref{affectation}, we represent it by boxplots.
 Closer there are to $1$, more different are the clusters.
 
In the present case, the two clusters are well defined and the clustering problem is quite easy, but see for example Fig. \ref{propDaily}, in a different clustering issue, 
which presents a model with affectations less well separated.

 \subsubsection{Clustering}
\begin{figure}[!ht]
\centering
\includegraphics[scale=0.24]{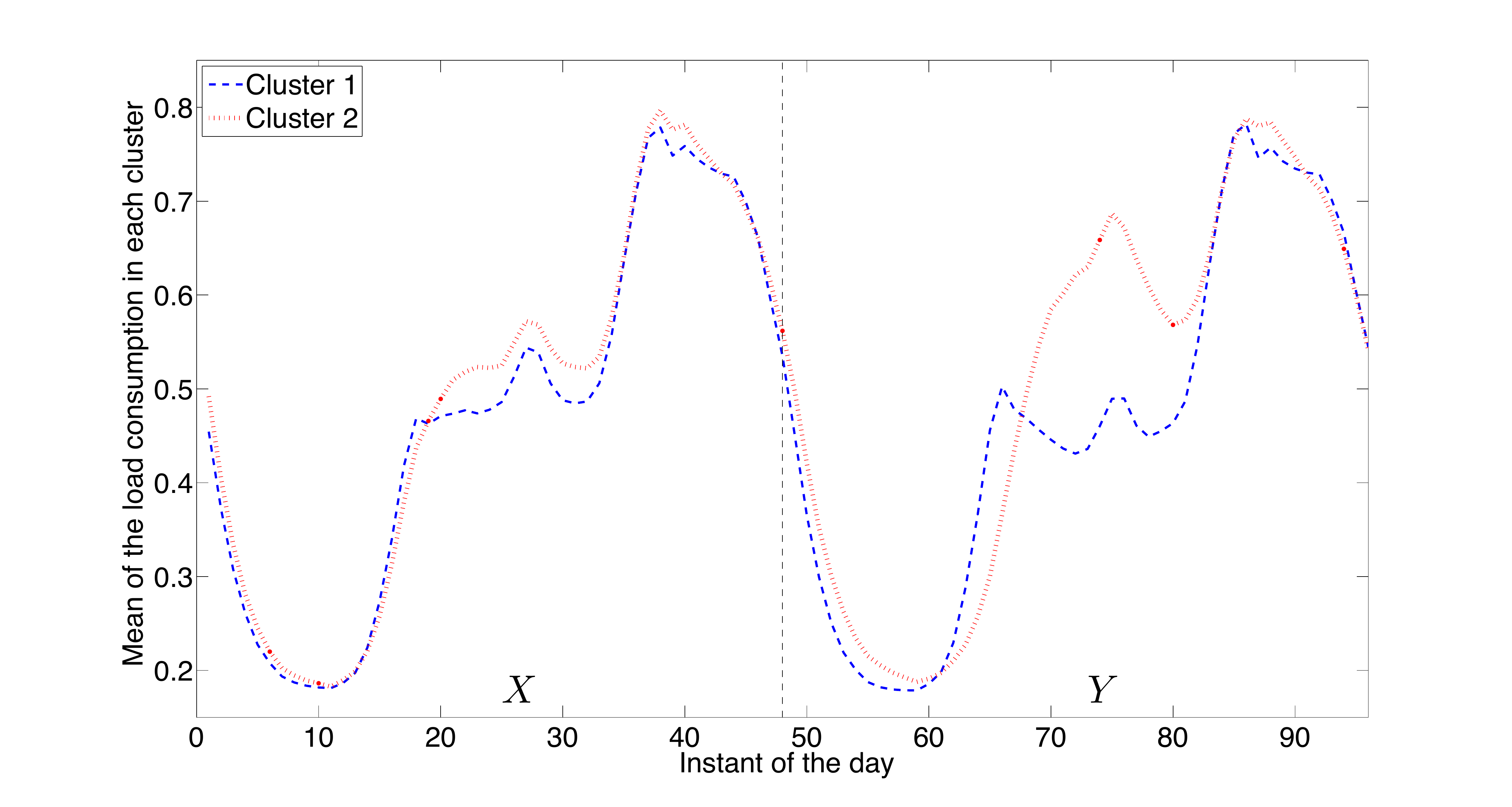} 
\caption{Clustering representation. Each curve is the mean in each cluster}
\label{courbesMoyennesPT1}
\end{figure}

 Now, we are able to try to interpret  each cluster.
 In Fig. \ref{courbesMoyennesPT1}, we represent the mean curves for each cluster.
 We can also use functional descriptive statistics (see \cite{Rainbow}).
 Because clusters are done on the reliance between a day and its eve, we represent the both days.
 
\subsubsection{Discussion}
 \label{discussion}
\begin{figure}[!ht]
    \centering
  \includegraphics[scale=0.24]{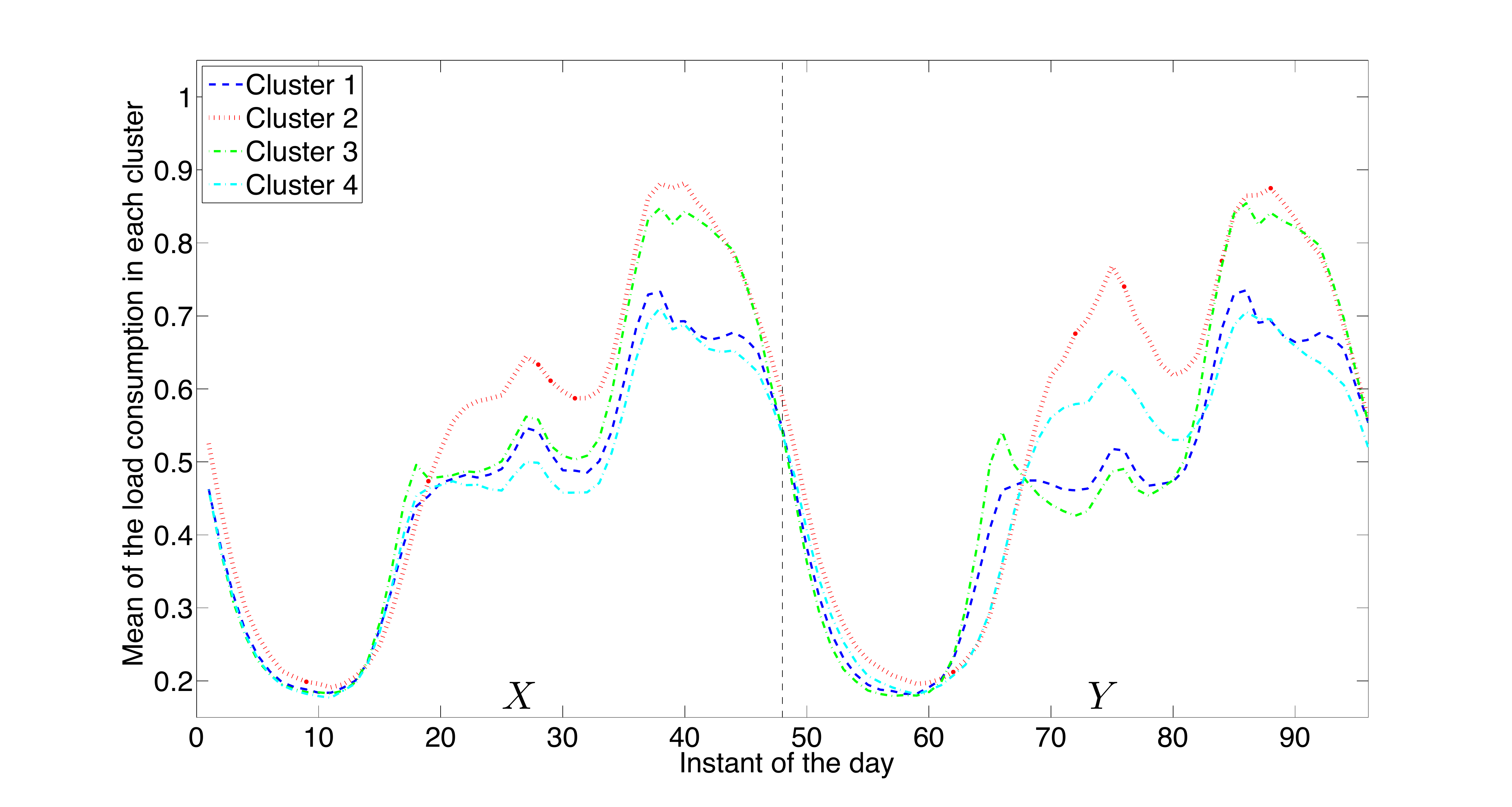}
  \caption{Clustering representation. Each curve is the mean in each cluster}
  \label{courbesMoyennesPT2}
 \end{figure}

 According to the preprocessing $2$, we cluster weekdays and weekend days.
 The same procedure done with the preprocessing $1$ shows the temperature influence.
 We construct four clusters, two of them being weekend days, and the two others are weekdays, differences made according to the temperature.
 In Fig. \ref{courbesMoyennesPT2}, we summarize clusters by the mean curves for this second model.
 
 In Table \ref{tableauJours}, we summarize the both models considered according to the day type.
 
 \begin{table}[!ht]
 \centering

 \begin{tabular}{|c|c|c|c|c|c|c|c|}
   \hline
   Interpretation& Mon. & Tue. & Wed. & Thur. & Fri. & Sat. & Sun. \\
   \hline
   week& 0.88 & 0.96& 0.94& 0.98& 0.96& 0&0 \\
   weekend& 0.12& 0.04& 0.06& 0.02& 0.04&1 &1 \\
   \hline
   week, low T.& 0.26 & 0.46 & 0.46 & 0.47 & 0.51 & 0 & 0 \\
   weekend, low T.& 0.1 & 0.02 & 0.03 & 0& 0 & 0.2 & 0.65 \\
   week, high T.& 0.64 & 0.52 & 0.5 & 0.53 & 0.45 &  0 &  0 \\
  weekend, high T. &0&0&0&0&0.04&0.79&0.35 \\
   \hline
 \end{tabular}
 \caption{For each model selected,
 we summarize the proportion of day type in each cluster, and interpret it, $T$ corresponding to the temperature.}

 \label{tableauJours}
 \end{table}
 
 According to Table \ref{tableauJours}, difference between both preprocessing is the temperature influence: when we center curves before projecting, we translate the whole curves,
 but when we remove the low frequency approximation, we skip the main features.
 Depending on the goal, each preprocessing could be interesting.
 
 \section{Clustering consumers}
 Another important approach considered in this paper is to cluster consumers. Before dealing with their daily consumption, in Section \ref{individual curves}, which is an hard problem because of 
 the irregularity of signals, we cluster consumers on mean days in Section \ref{MeanDays}.
 \subsection{Cluster consumers on mean days}
 \label{MeanDays}
 Define $\tilde{\mathfrak{z}}_{i,d}$ the mean signal over all days $d$ for the residential $i$, over all days $d \in \{1,\ldots,7\}$.
 Then we consider couples $(\tilde{\mathfrak{z}}_{i,d},\tilde{\mathfrak{z}}_{i,d+1})_{1 \leq i \leq n}$.
 
 If we look on the model collection constructed by our procedure for $\mathcal{K} = \{1,\ldots,8\}$, we always select models with only one component, for every days $d$.
 Nevertheless, if we force the model to have several clusters, restricting $\mathcal{K}$ to $\mathcal{K}^{'} = \{2,\ldots,8\}$, we get some interesting informations.
 All results get here are done for preprocessing $2$.
 
 For weekdays couples, Monday/Tuesday, Tuesdays/Wednesday, Wednesday/Thursday, Thursday/Friday, we select models with two clusters, with same means and same covariance matrices:
 the model with one component is needed. The only difference on load consumption is on the mean comportment. It is relevant with clusterings obtained in Section \ref{ClusterDays}.

%  \begin{figure}[ht!]
%  \centering
%   \includegraphics[scale=0.15]{FridaySaturday}
%   \caption{Means of Friday and Saturday in each cluster}
%  \label{FridaySaturday}
%  \end{figure}

\begin{figure}[ht!]
 \centering
  \includegraphics [scale=.21]{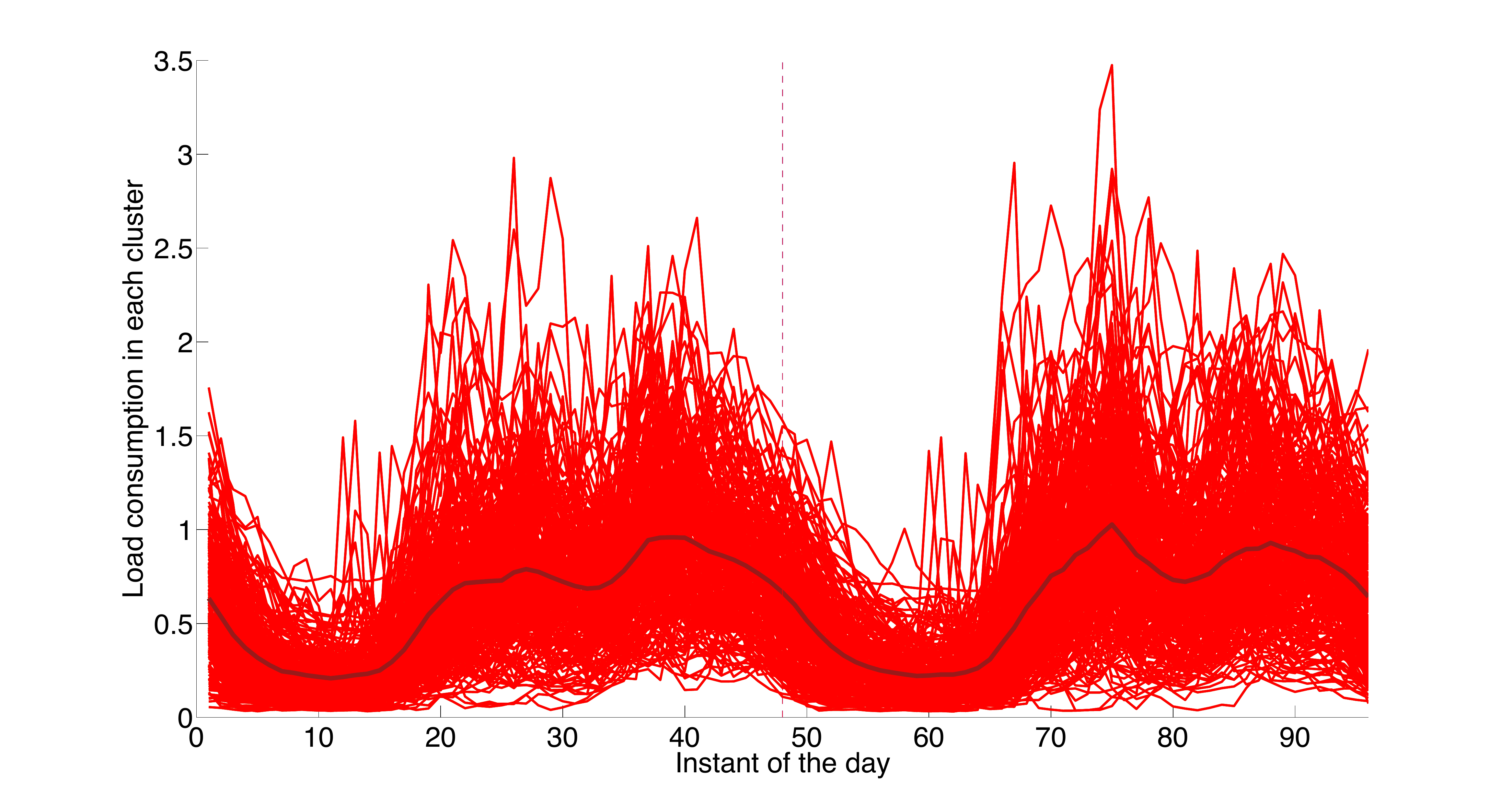}
  \includegraphics [scale=.21]{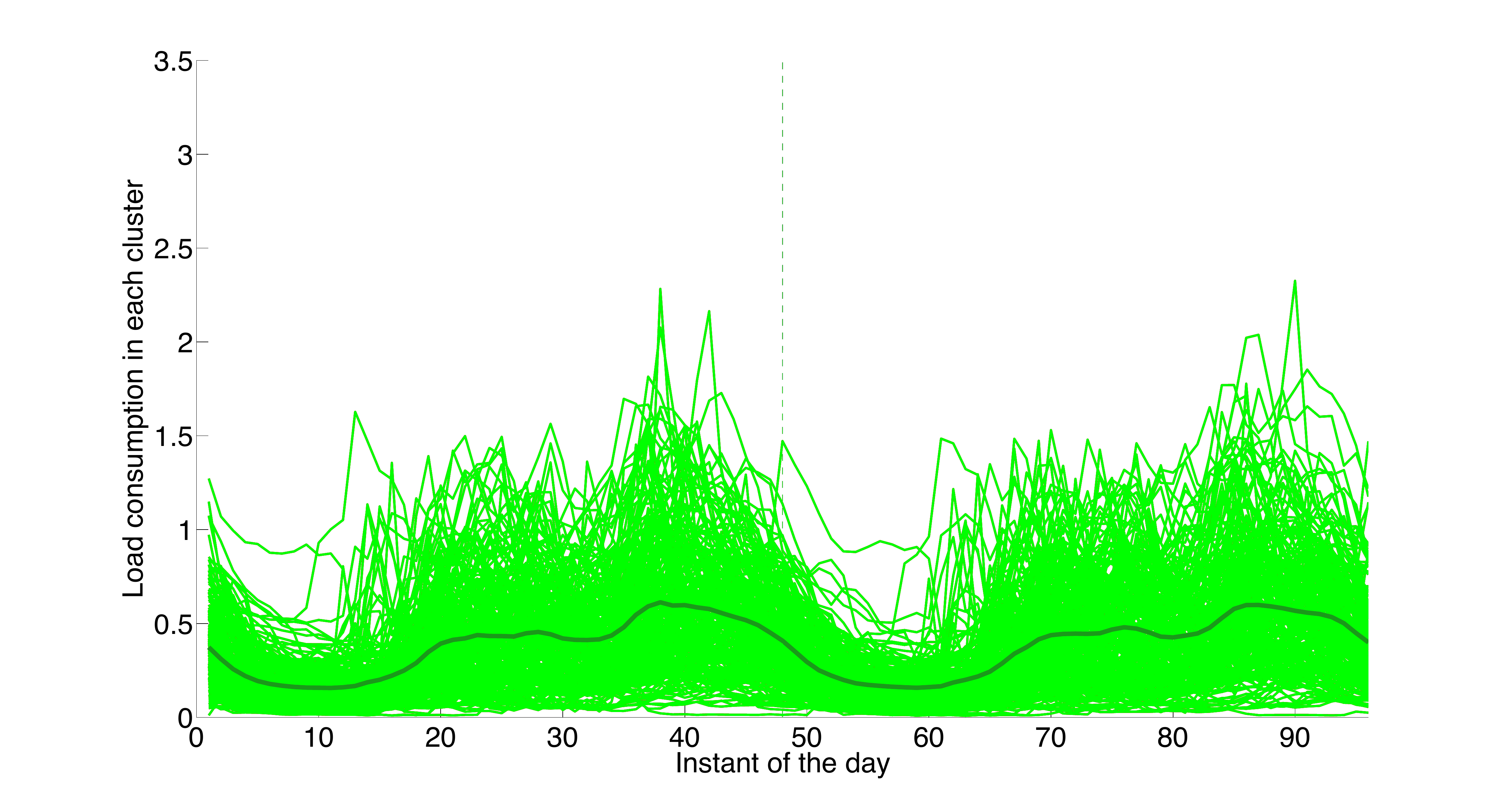}
  \caption{Saturday and Sunday load consumption in each cluster.}
  \label{SaturdaySunday}
 \end{figure}

 We focus here on Saturday/Sunday, for which there are different interesting clusters, see Fig. \ref{SaturdaySunday}.
Remark that we cannot summarize a cluster to its mean because of the high variability.
The main differences between those two clusters are on differences between lunch time and afternoon, and on the Sunday morning.
Notice that the big variability over the two days is not explained by our model, for which the variability is small, explaining the noise for the reliance between a day and its eve.
 
 On Sunday/Monday, we get also three different clusters.
 Even if we identify differences on the shape, the main difference is still on the mean consumption.
 On Friday/Saturday, we see differences between people who have the same consumptions, and people who have a really different comportment.
%  \begin{figure}[ht!]
%  \centering
%   \includegraphics[scale=0.15]{DimancheLundi}
%   \caption{Means of Friday and Saturday in each cluster}
%  \label{DimancheLundi}
%  \end{figure}

 However, because the selected model is again with one component, we think that consider the mean over days of each consumer cancel interesting effects, as the temperature and seasonal influence.

 \subsection{Cluster consumers on individual curves}
 \label{individual curves}

 One major objective of this work is individual curves clustering. As already pointed in the introduction, this is a very challenging task that has various applications for smart grid management, going from  demand response programs to energy reduction recommendations or household segmentation. We consider the complex situation where an electricity provider as access to very few recent measurements -2 days here- of each individual customers but need to classify them anyway. That could happen in a competitive electricity market where customers can change their electricity provider at any time without providing their past consumption curves.

 First, we focus on two successive days of electricity consumption measurements - Tuesday and Wednesday in winter: January 5th and 6th 2010- for our 487 residential customers. Note that we choose week days in winter following our experience on electricity data, as those days often bring important information about residential electricity usage (electrical heating, intra-day cycle, tariff effect...).
 
  \subsubsection{Selected mixture models}
 \label{WeekdaysWinter}
\begin{figure}[!ht]
     \centering
  \includegraphics[scale=0.26]{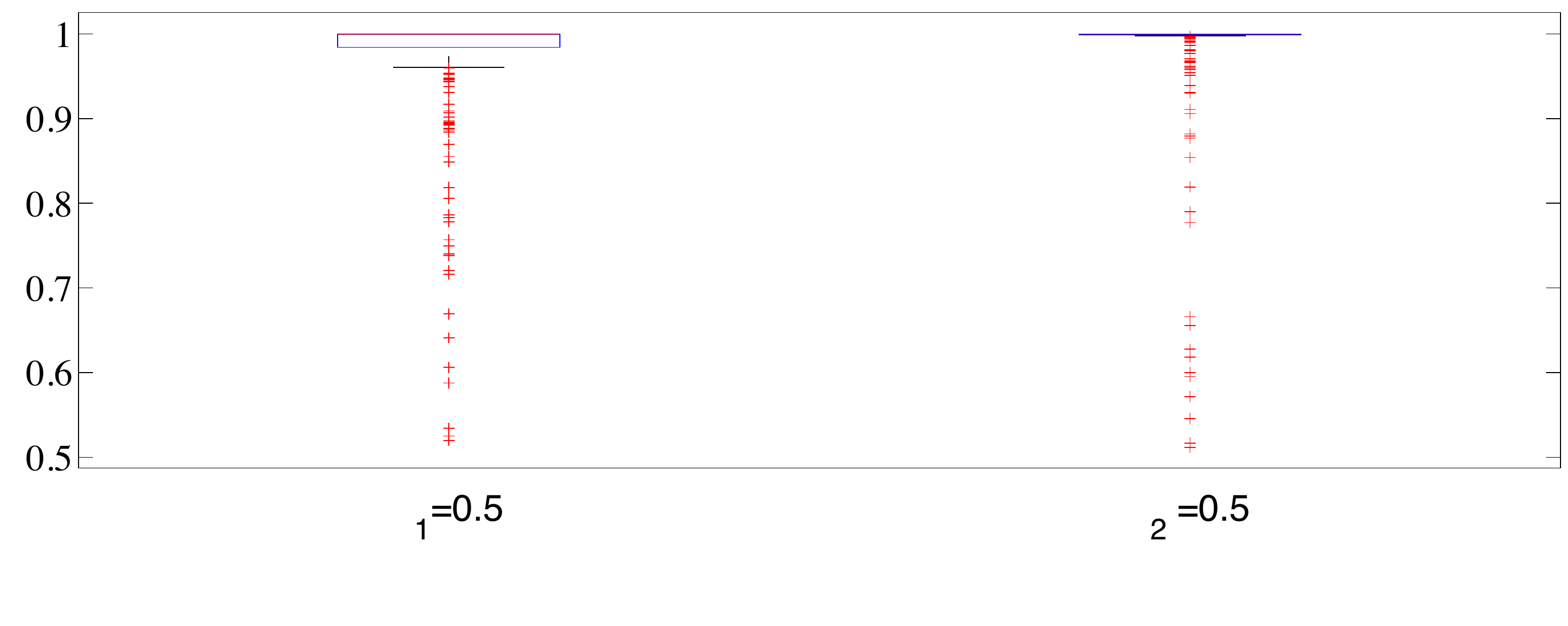}
  \includegraphics[scale=0.25]{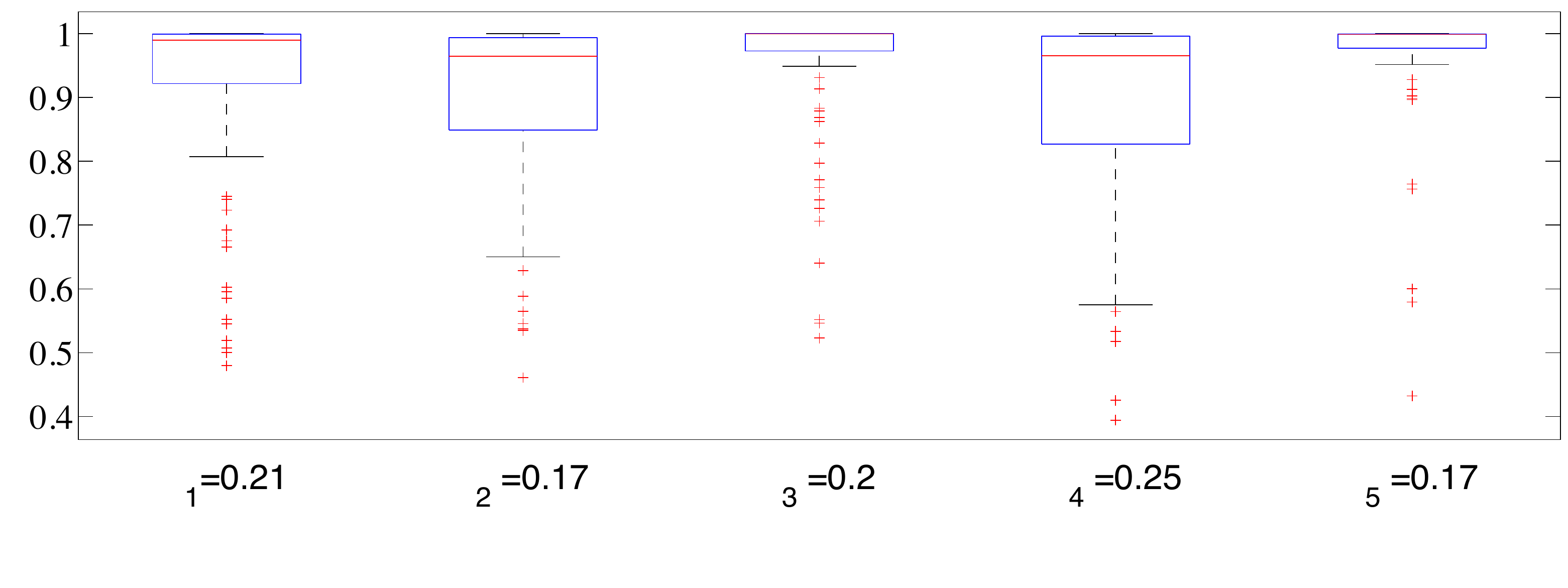}
  \caption{Proportions in each cluster for models constructed by our procedure}
  \label{propDaily}
 \end{figure}

 We apply the model-based clustering approach presented in Section \ref{Method} for preprocessing $1$ and obtain two models minimizing the penalized log-likelihood corresponding to 2 and 5 clusters.
 Although these two classifications are based on auto-regression mixture model, we are able to analyze and interpret clusters in terms of consumption profiles and provide bellow a physical interpretation for the classes. 
 In Fig. \ref{propDaily}, we plot the proportions in each cluster for both models constructed by our procedure.
 The first remark is that this issue is harder than the one in Section \ref{ClusterDays}. Nevertheless, even if there are a lot of outliers, for the model M1, a lot of affectations are well-separated. It is obviously less clear with the model M2, with $5$ components.
 
\begin{figure}[!ht]
 \begin{minipage}[c]{.48\linewidth} 
 \centering
  \includegraphics[scale=0.2]{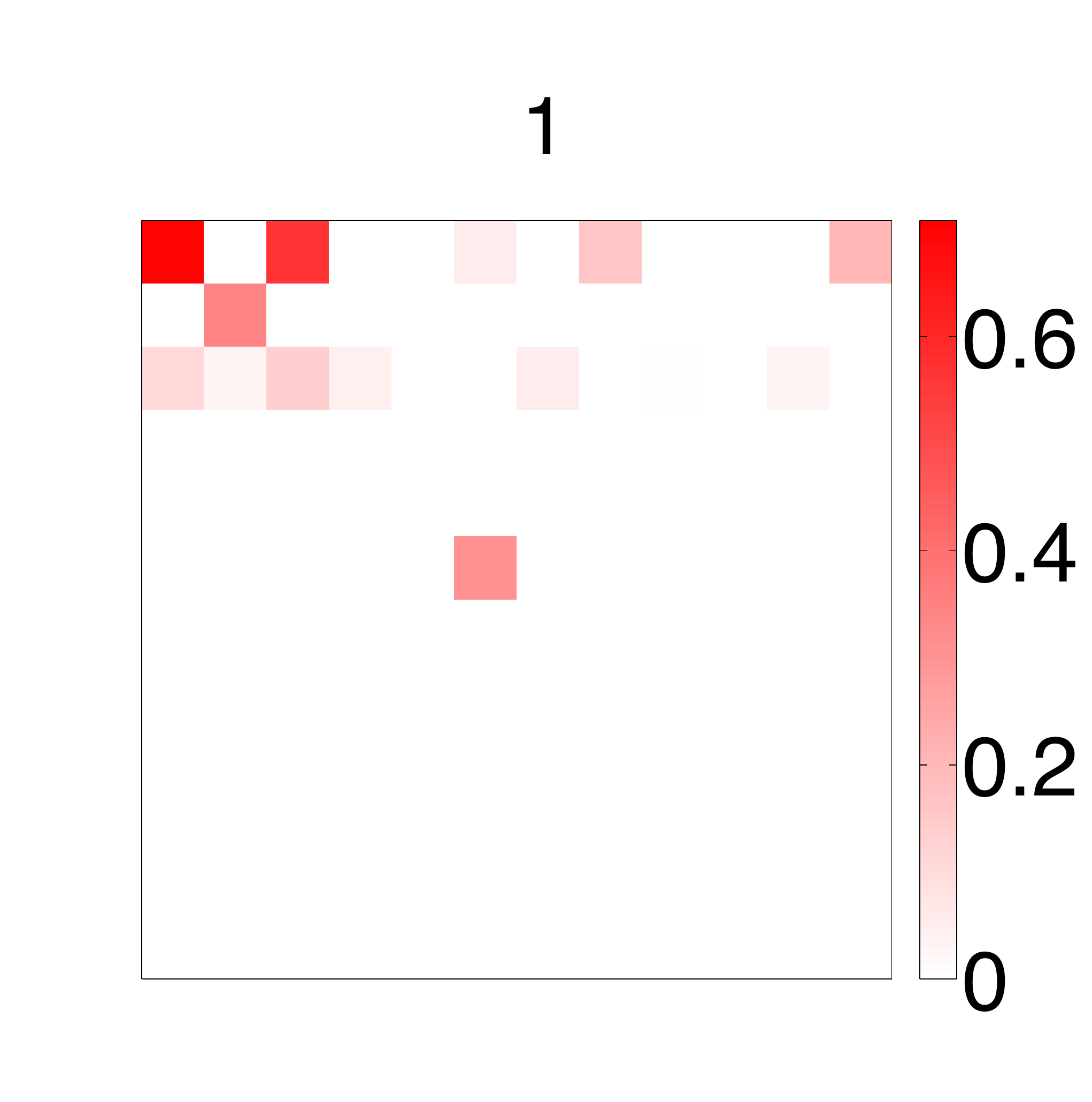}
 \end{minipage}
 \begin{minipage}[c]{.48\linewidth} 
 \centering
 \includegraphics[scale=0.2]{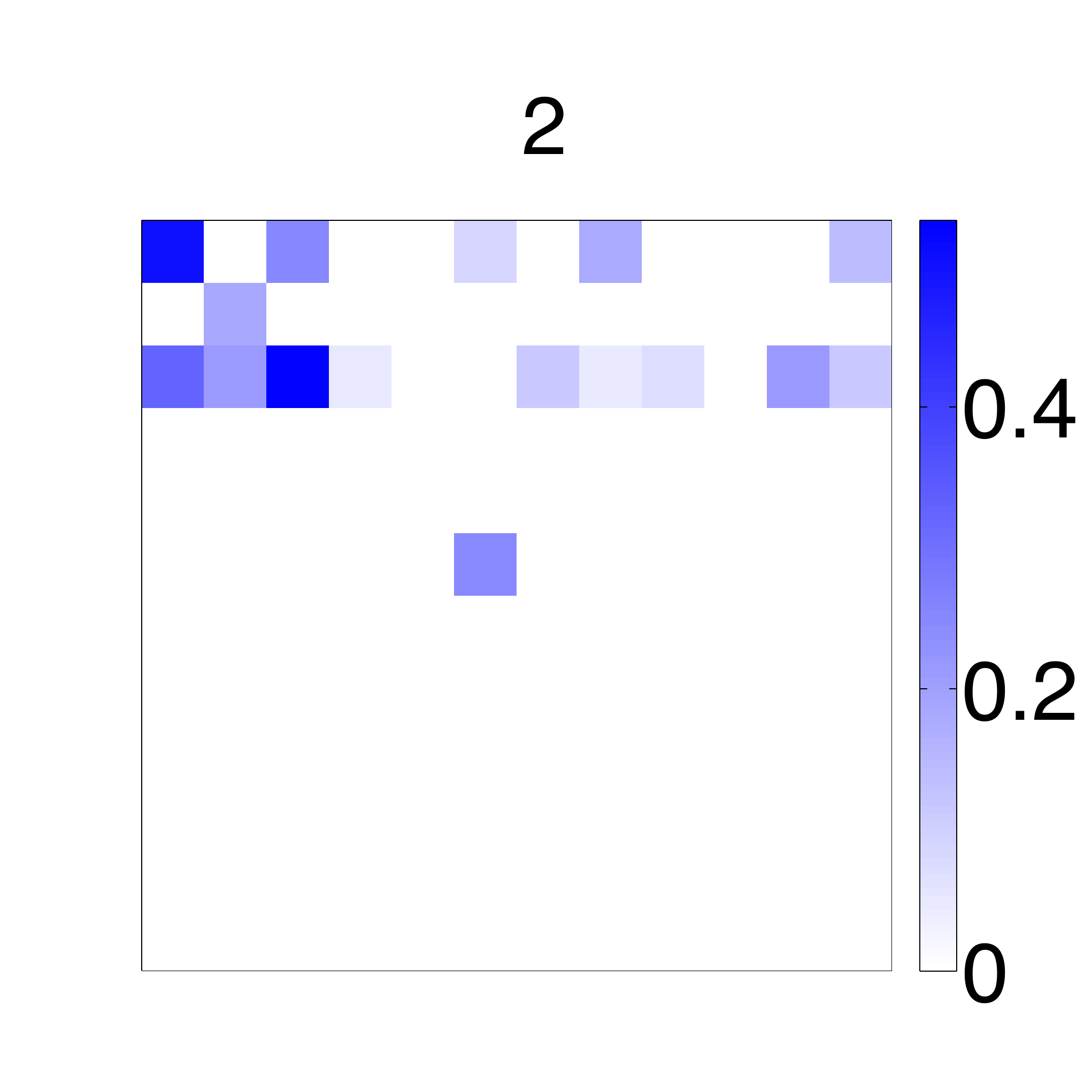}
 \end{minipage}
    \caption{Regression matrix in each cluster for the model with 2 clusters}
  \label{betaDaily}
 \end{figure}

 In Fig. \ref{betaDaily}, we plot the regression matrix to highlight differences between clusters. Indeed, those two matrices are different, for example more variables are needed to describe the cluster $2$.
 
 \subsubsection{Analyzing clusters using classical profiling features}
\begin{figure}[!ht]
    \centering
  \includegraphics[scale=0.6,trim = 0cm 10cm 1cm 2cm, clip]{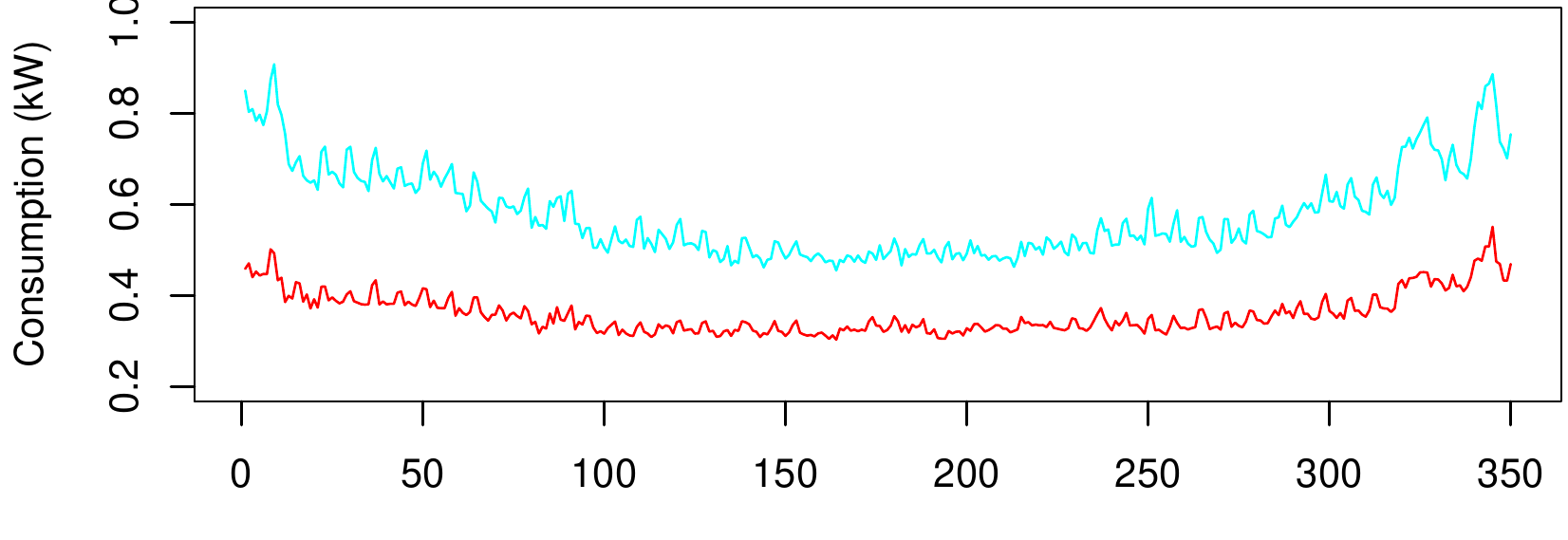}
  \includegraphics[scale=0.6,trim = 0cm 0.5cm 1cm 10.5cm, clip]{dailyMean_winter.pdf}
  \caption{Daily mean consumptions of the cluster centres along the year for 2 (top) and 5 clusters (bottom)}
  \label{dailyMean_winter}
\end{figure}

We first represent the main features of the cluster  centres (the mean of all individual curves in a cluster). Fig. \ref{dailyMean_winter} represents daily mean consumptions of these centres along the year. We clearly see that the two classifications separate customers that have different mean level of consumption (small, -middle- and big residential consumers) and different ratio winter/summer consumption probably due to the amount of electrical heating among house usage. Let recall that the model based clustering is done on centered data and that the mean level discrimination is not straightforward. Schematically, the 2 clusters classification seems to separate big customers with electrical heating from small customers with few electrical heating. Whereas the 5 clusters classification separates the small customers with few electrical heating but peaks in consumption (flat curve with peaks, probably due to auxiliary heating when temperature is very low) and middle customers with electrical heating. The two clusters in the middle customers population don't present any visible differences on this figure. The two big customers clusters have a different ratio winter/summer probably due to a difference in terms of electrical heating usage.

\begin{figure}[!ht]
\begin{center}
\begin{minipage}{0.48 \linewidth}
\begin{center}
\includegraphics[scale=0.4,trim = 1cm 0cm 1cm 2cm, clip]{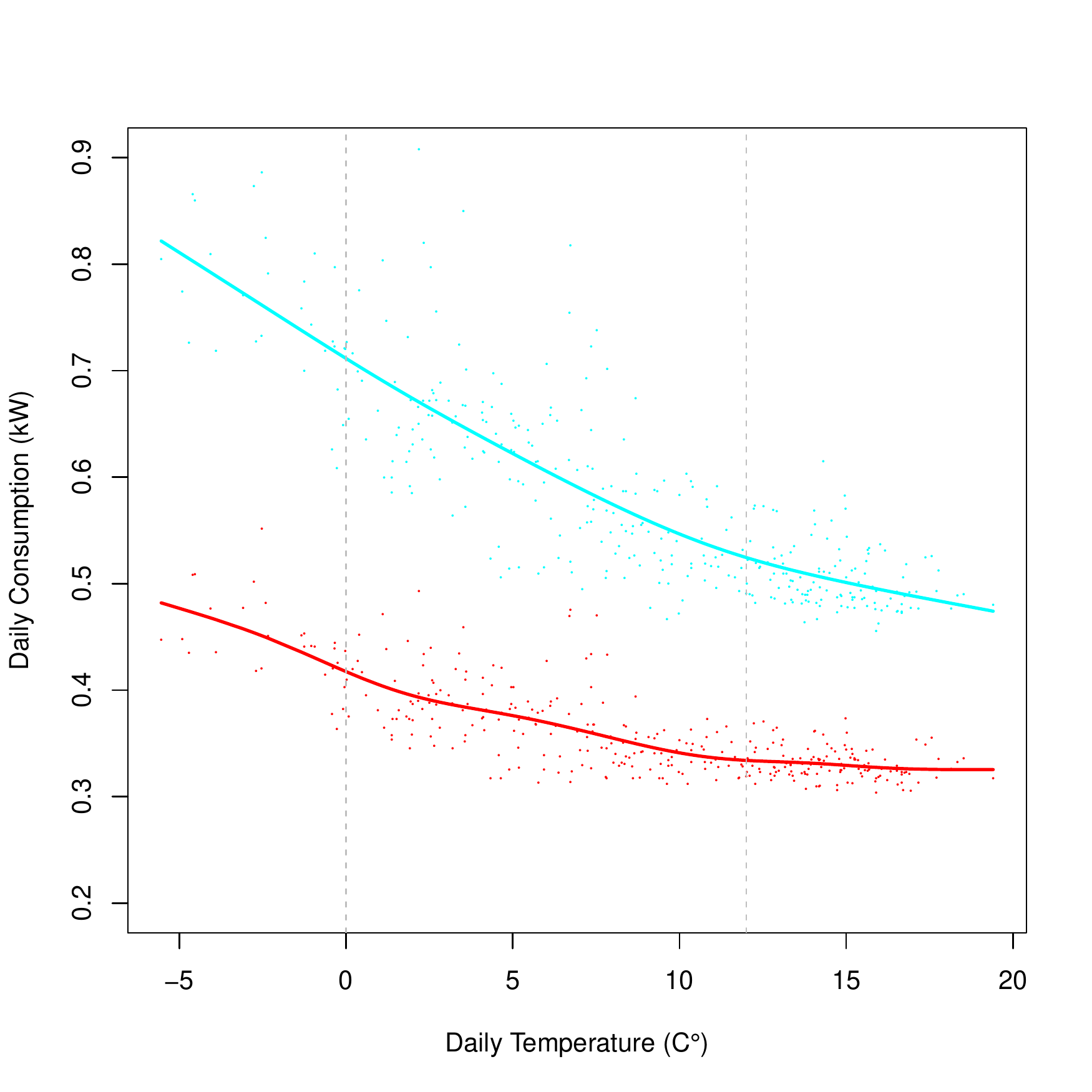}
\end{center}
\end{minipage}
\begin{minipage}{0.48 \linewidth}
\begin{center}
\includegraphics[scale=0.4,trim = 1cm 0cm 1cm 2cm, clip]{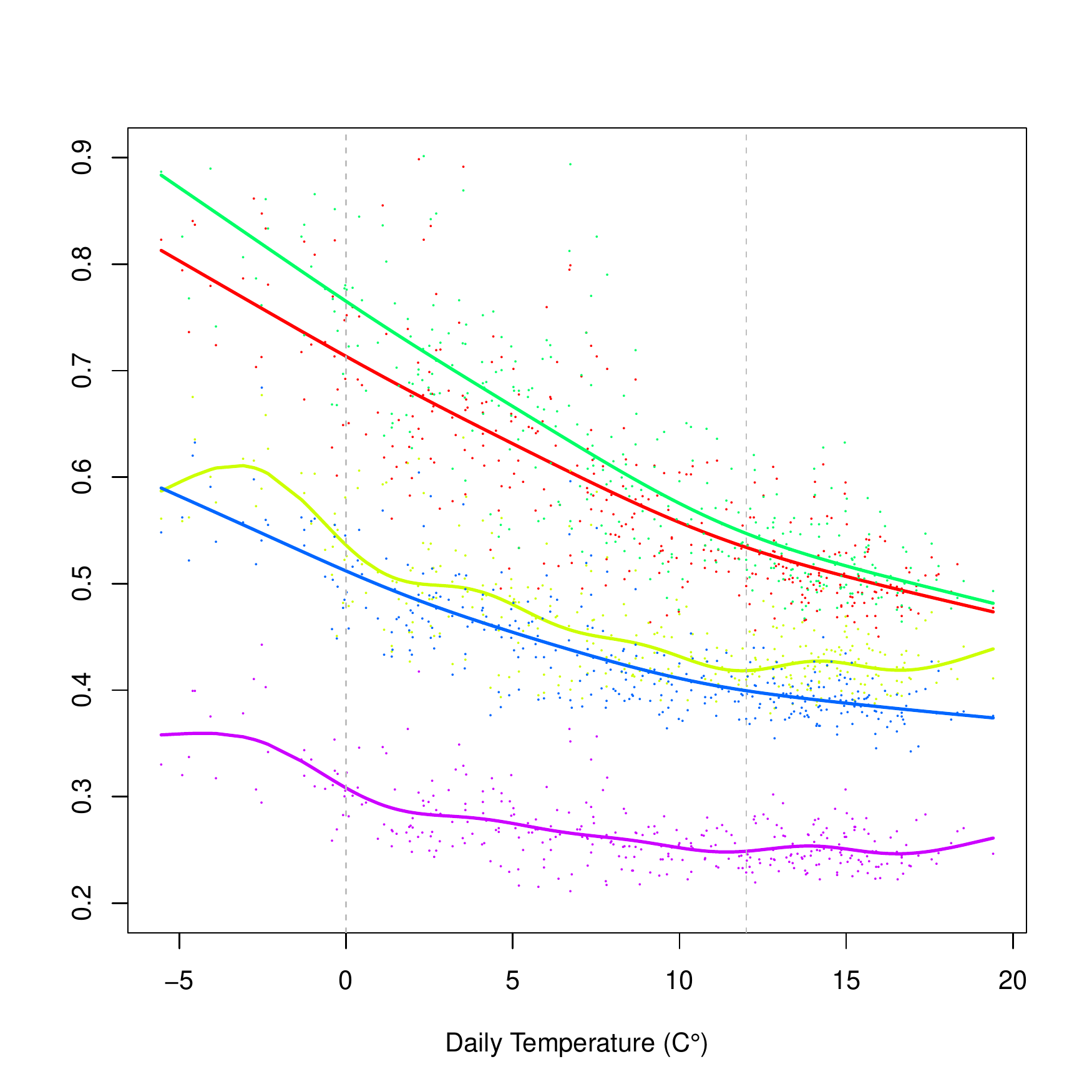}
\end{center}
\end{minipage}
\end{center}

\caption{Daily mean consumptions of the cluster centres in function of the daily mean temperature for 2 (on the left) and 5 clusters (on the right)} 
\label{temperature_conso_winter}
\end{figure}

This analysis is confirmed in Fig. \ref{temperature_conso_winter} where we represent the daily mean consumptions of the cluster centres in function of the daily mean temperature for the two classifications. Points correspond to the mean consumption of one day and smooth curves are obtained with P-spline regression. We observe that for all classes, the temperature effect due to electrical heating starts at around 12 \textcelsius and that the different clusters have various heating profiles. The 2 clusters classification profiles confirm the observation of Fig. \ref{dailyMean_winter} that the population is divided into small and big customers with electrical heating. Concerning the 5 clusters classification, we clearly see on the small customer profile (purple points/curves) an inflexion at around 0 \textcelsius -this inflexion is also observed in the small customer cluster of the 2 clusters classification- corresponding to e.g. an auxiliary heating device effect or at least an increase of consumption of the house for very low temperature. This is also what distinguishes the 2 middle customers classes (blue and green points/curves). The two big customers' clusters have similar heating profile, except that the green cluster correspond to higher electrical heating usage.

% \begin{figure}[!ht]
% \begin{minipage}[c]{.48\linewidth} 
% \centering
%   \includegraphics[width=4cm,height=4cm]{temperature_conso_winter1.pdf}
% \end{minipage}
% \begin{minipage}[c]{.48\linewidth} 
% \centering
%   \includegraphics[width=4cm,height=4cm]{temperature_conso_winter2.pdf}
% \end{minipage}
%   \caption{Daily mean consumptions of the cluster centres in function of the daily mean temperature for 2 and 5 clusters}
%      \label{temperature_conso_winter}
% \end{figure}

\begin{figure}[!ht]
    \centering
  \includegraphics[scale=0.55,trim = 0cm 10cm 1cm 2cm, clip]{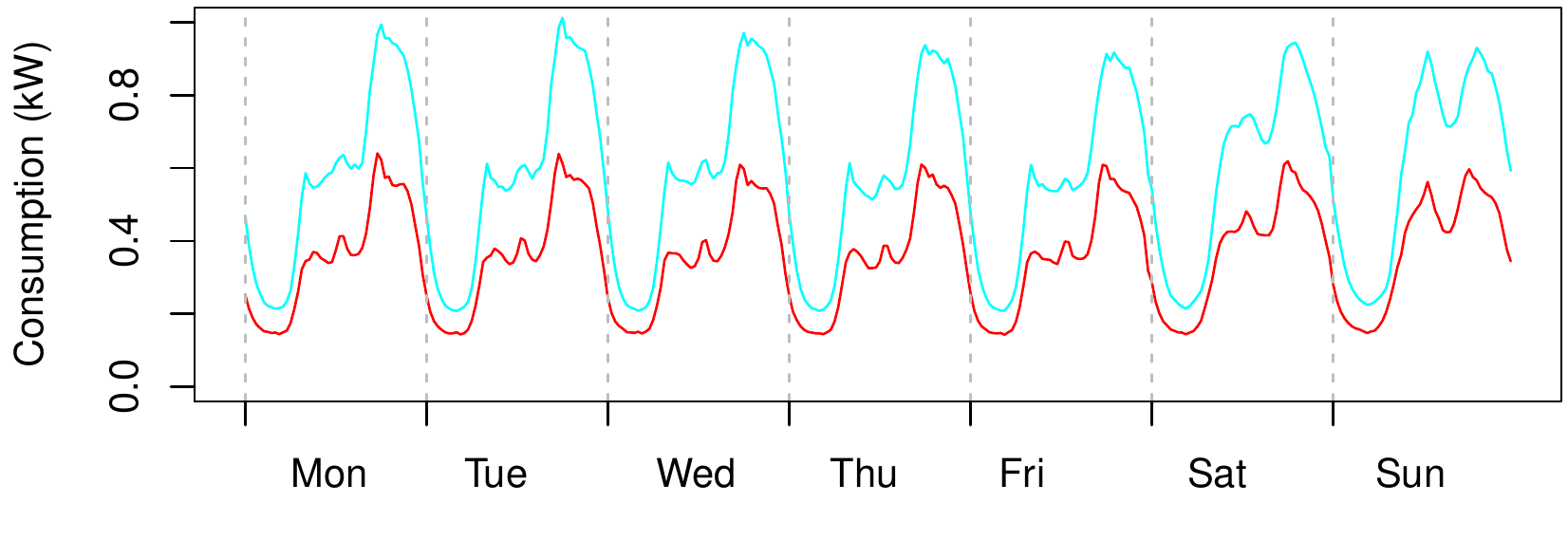}
  \includegraphics[scale=0.55,trim = 0cm 1.5cm 1cm 10.5cm, clip]{WeeklyProfile_winter.pdf}
  \caption{Average (over time) week of consumption for each centre of the two classifications (2 clusters on the top and 5 on the bottom)}
  \label{WeeklyProfile_winter}
\end{figure}

Another interesting observation concern the weekly and daily profiles of the centres. We represent on Fig. \ref{WeeklyProfile_winter} an average (over time) week of consumption for each centre of the two classifications. For the 2 clusters classification, we see again the difference in average consumption between the big customer cluster profile and the small customer one. They have similar shapes but the difference day/night and peak loads (at around 8h, 13h and 18h for weekdays), are more marked. For the 5 clusters curves, even if the weekly profiles are quite similar (no major difference in the week days/week ends profiles in each clusters), the daily shapes exhibit more differences. The day/night ratio could be very different as well as small variation along the day, probably related to different tariff options (see \cite{Cer_a} for a description of the tariffs).

\subsubsection{Cross analysis using survey data}
\begin{figure}[!ht]
    \centering
  \includegraphics[scale=0.4]{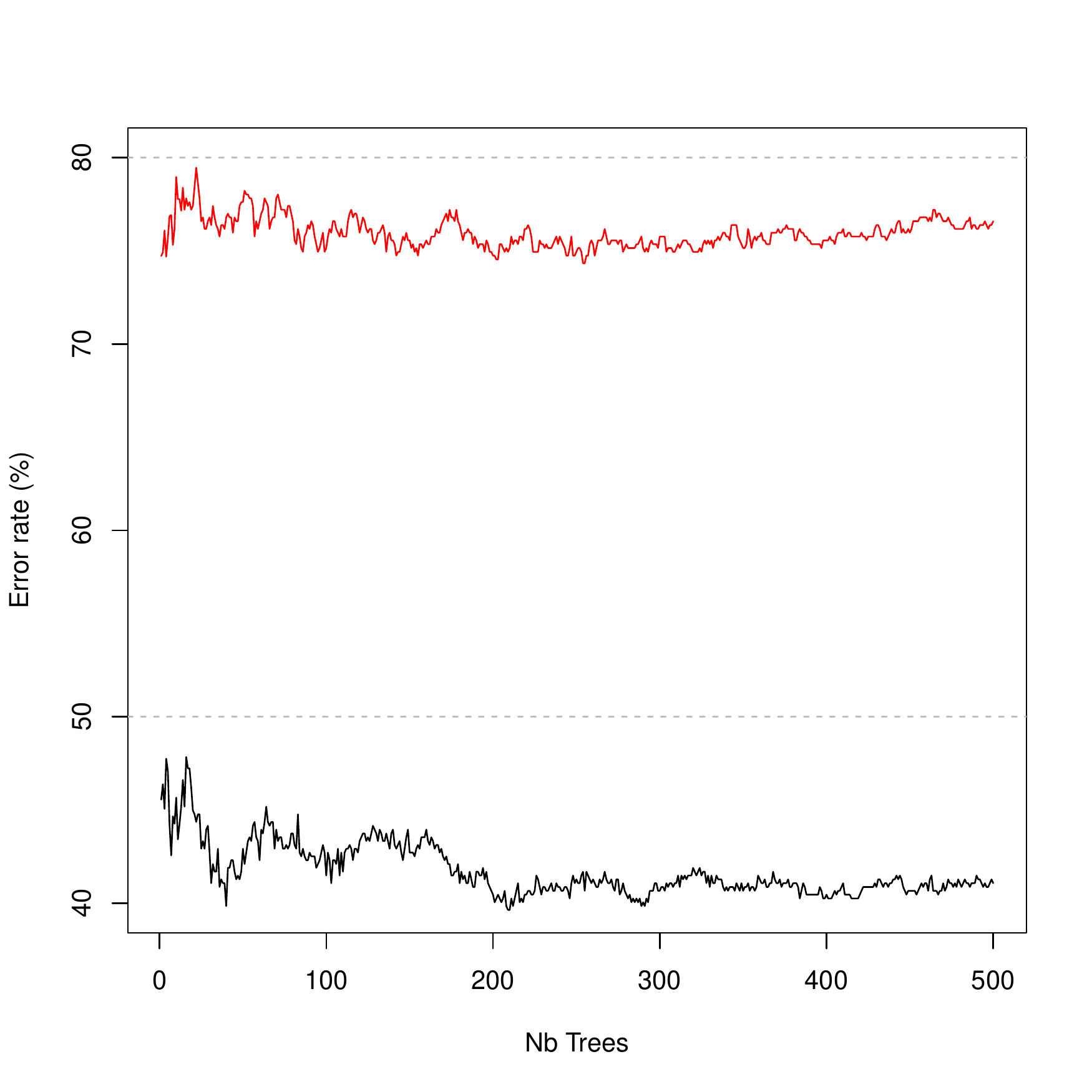}
  \caption{Out of bag error of the random forest classifiers in function of the number of trees}
  \label{rf_error_rate}
\end{figure}

To enrich this analysis, we also consider extra information providing in a survey realized by the Irish Commission for Energy Regulation. We summarize this large amount of information into 10 variables:
\textit{ResidentialTariffallocation} corresponds to the tariff option (see \cite{Cer_a}, \cite{Cer_b}), \textit{Socialclass}: AB, C1, C2, F in UK demographic classifications, \textit{Ownership} whether or not a customer owns his/her house, \textit{ResidentialStimulusallocation} the stimulus sends to the customer (see \cite{Cer_a}, \cite{Cer_b}), \textit{Built.Year} the year of construction of the building, \textit{Heat.Home} and \textit{Heat.Water} electrical or not, \textit{Windows.Doubleglazed} the proportion of double gazed window in the house (none, quarter, half, 3 quarters, all), \textit{Home.Appliance.White.goods} the number of white goods/major appliances of the household. To relate our clusters to those variables we consider the classification problem consisting in recovering model based classes with a random forest classifier. Random forest introduced in \cite{Breiman2001} is a well known and tested non-parametric classification method that has the advantage to be quite automatic and easy to calibrate. In addition, it provides a nice summary of the previous covariate in terms of their importance for classification. On the Fig. \ref{rf_error_rate} we represent the out of bag error of the random forest classifiers in function of the number of trees (one major parameter of the random forest classifier) for the two clusterings.
That corresponds to a good estimate of what could be the classification error on a independent dataset. We have observed that, choosing a sufficiently large number of trees for the forest (300), the classification error rate attains 40\% in the 2 clusters case and 75\% in the 5 classes case which has to be compared to a random classifier who has respectively a 50\% and 80\% error rate. That means that the 10 previous covariates provide few but some information about the clusters. 

Quantifying the importance of each survey covariate in the classifications we observe that in both cases, 
\textit{Home.Appliance.White.goods} and \textit{Socialclass} play a major role. That could be explained as those covariates could discriminate small and big customers.
Another interesting point is that in the 5 clusters classification, the variable \textit{Built.Year} plays an important role which is probably related to different
heating profiles explained by different isolation standards. That could explain the two big customers clusters. Then come the tariffs options which, in the 5 clusters case, could explain the different daily shapes of the Fig. \ref{WeeklyProfile_winter}.

% 
% \begin{figure}[!ht]
%     \centering
%   \includegraphics[scale=0.45]{rf_importance.pdf}
%   \caption{Variable importance for the 2 classifications}
%   \label{rf_importance}
% \end{figure}

It is noteworthy that these two classifications provide clusters that exhibit those very nice physical interpretation, considering that we only use two days of consumption in winter for each customer.

\subsubsection{Using model on one-day shifted data}
To highlight advantages of our procedure, we compare prediction for the consumption of Thursday, 7th January, 2010. Indeed, even if the method is not designed for forecasting purposes,
we want to show that model-based clustering is an interesting tool also for prediction. We will
compare linear models, estimated on couples Tuesday, 5th January and Wednesday, 6th January, and we will predict Thursday from Wednesday. 
This is suggested by the clustering get in Section \ref{ClusterDays}, showing that transitions between weekdays are similar.
We then compare the following models.
First, the most common is the linear model, without clustering.
The second model is the first constructed by our procedure, described before, with $2$ components. Moreover, we could use the clustering get by the models constructed by our procedure, but estimate parameters without variable selection, using full linear model in each component. We restrict here our study to one model to narrow 
the analysis, but everything is also computable with the models with $5$ clusters.

% Finally, we compute the $k$-means algorithm, with $k=2$, to compute other clustering.
% Denote that this last clustering is done on Tuesday datasets, projected into the Haar basis, according to the preprocessing $1$.
% Then, we will compare the use of clustering on one days and of clustering on transitions for prediction goal.

For each consumer $i$, for each prediction procedure, we compute two prediction errors: the RMSE on Thursday prediction, and the RMSE of Wednesday prediction.
Remind that RMSE, for a consumer $i$, is defined by
\begin{align*}
 RMSE(i) &= \sqrt{\frac{1}{48} \sum_{t=1}^{48} (\hat{z}_{i,t}-z_{i,t})^2}.
%  MAPE(i) &= \frac{1}{48} \sum_{t=1}^{48} \left| \frac{z_{i,t}-\hat{z}_{i,t}}{z_{i,t}} \right| .\\
\end{align*}

\begin{figure}[!ht]
\centering
 \includegraphics[scale=0.2]{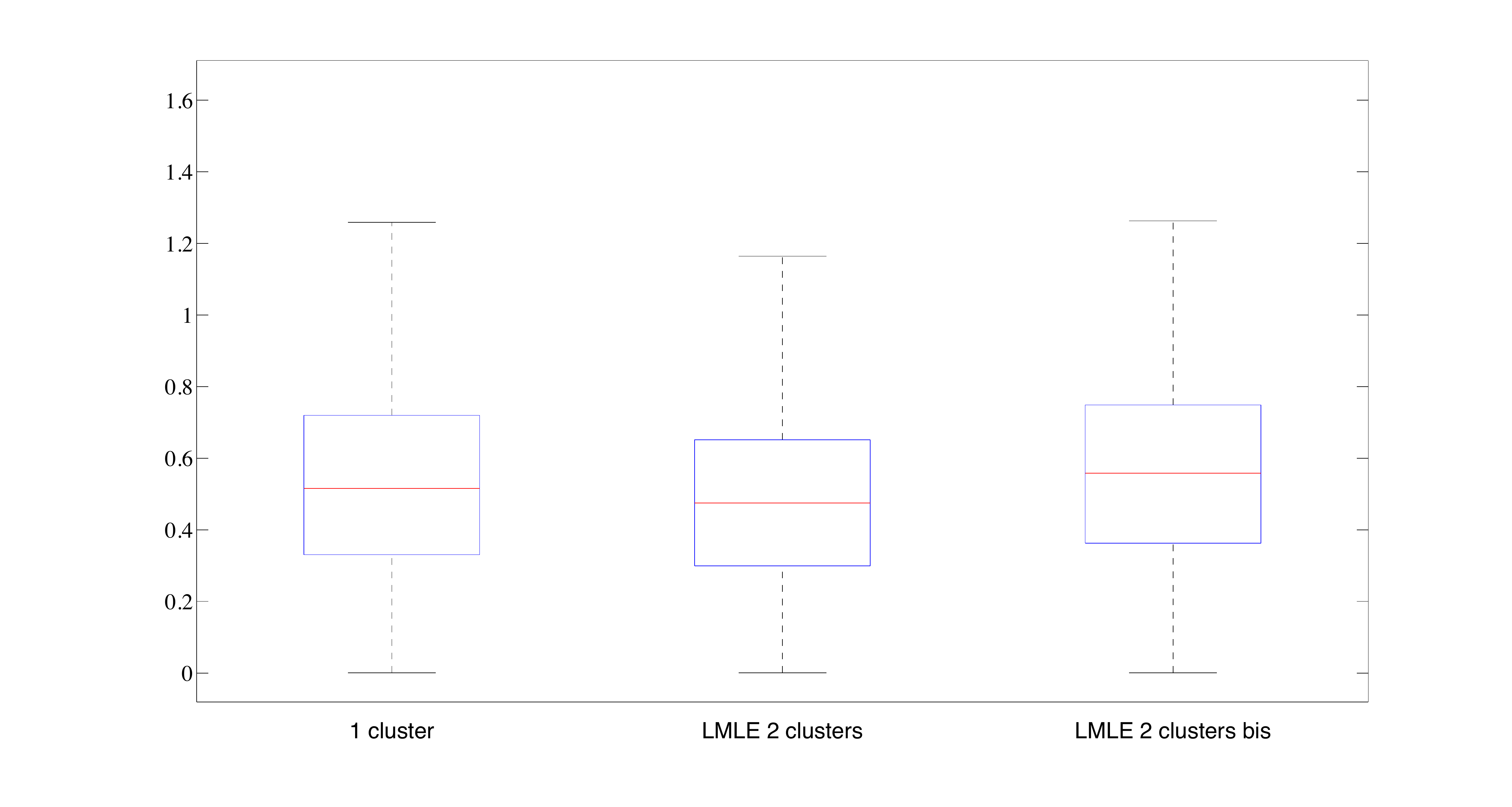}
 \caption{RMSE on Thursday prediction for each procedure over all consumers}
\end{figure}
%  \begin{figure}[!ht]
%   \centering
% %  \includegraphics[scale=0.2]{MAPEm1}
% %  \caption{MAPE for prediction constructed by each procedure over all consumers.} 
%   \includegraphics[scale=0.15]{ratioRMSE}
%   \caption{Ratio between RMSE  predictions constructed by each procedure over all consumers.}
% \end{figure}

We remark that if RMSE are almost the same for the three different models, the one estimated by our procedure leads to smaller median and interquartile range.
For the three considered models, the median of the RMSE on Wednesday prediction (learning sample) and the RMSE on Thursday prediction (test sample) are close to each other, which means that the clustering remains good, even for one-day shifted data, of course as long as we remain in the class of working days, according to Section \ref{ClusterDays}.
To highlight this remark, we also compute our procedure on couple Wednesday/Thursday. 
Then, we select three different models, and involved clusterings are quite similar to clusterings given by models in Section \ref{WeekdaysWinter}.
%If we compute our procedure on the sample Wednesday-Thursday, we select three different models, embedded.
%In addition, the two clusters given by the model described in Section  and  two clusters coming from the model fitted on this dataset are quite similar.

\subsubsection{Remarks on similar analyses}
 
Alternatively, we make the same analysis on two successive weekdays of electricity consumption measurement in summer.
We obtain three models, corresponding to $2$, $3$ and $5$ clusters respectively.
We compute, as in the Subsection \ref{WeekdaysWinter}, daily mean consumptions of the cluster center along the year, and in function of the daily mean temperature.
The main difference is about the inflexion at around 0 \textcelsius. Because clustering is done for summer days, we do not distinguish cold effects.
Moreover there are no cooling effects.
We could remark again that clusterings are hierarchical, but different from those get in the winter study, as we expected.
\begin{figure}[!ht]
\centering
 \includegraphics[scale=0.3,trim = 0cm 0cm 1cm 2cm, clip]{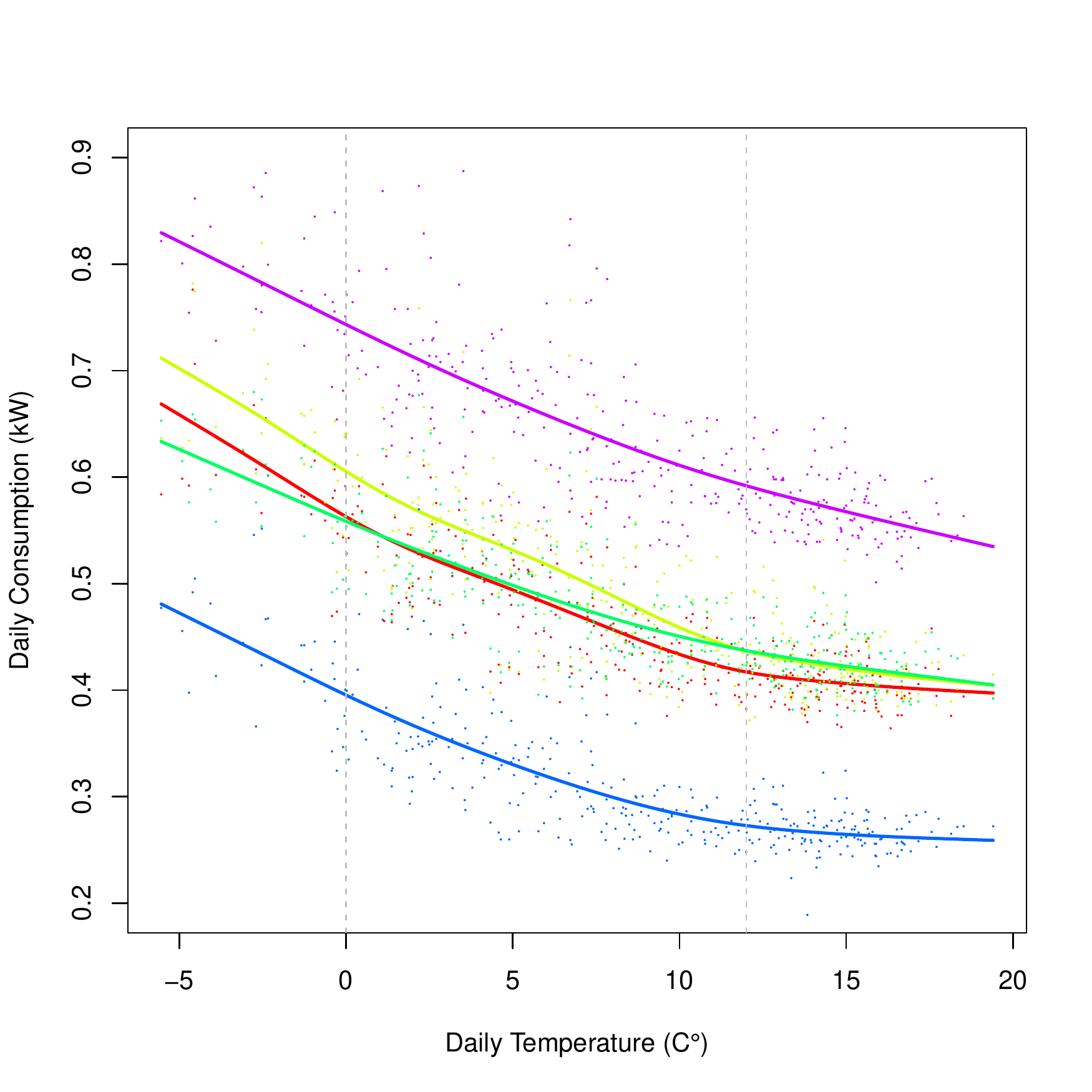}
 \caption{Daily mean consumptions of the cluster centres in function of the daily mean temperature for 5 clusters, clustering done by observing Thursday and Wednesday in summer}
\end{figure}

We also study two successive weekend days of electricity consumption, in winter and in summer.
We recognize different clusters, depending on behavior of consumers.
We work with Friday/Saturday couples.
The main thing we observe in summer is a cluster with no-consumption, consumers who leave their home.
It could be useful to predict the Sunday consumption, but no more general for other weekend.

\begin{figure}[!ht]
\centering
 \includegraphics[scale=0.15]{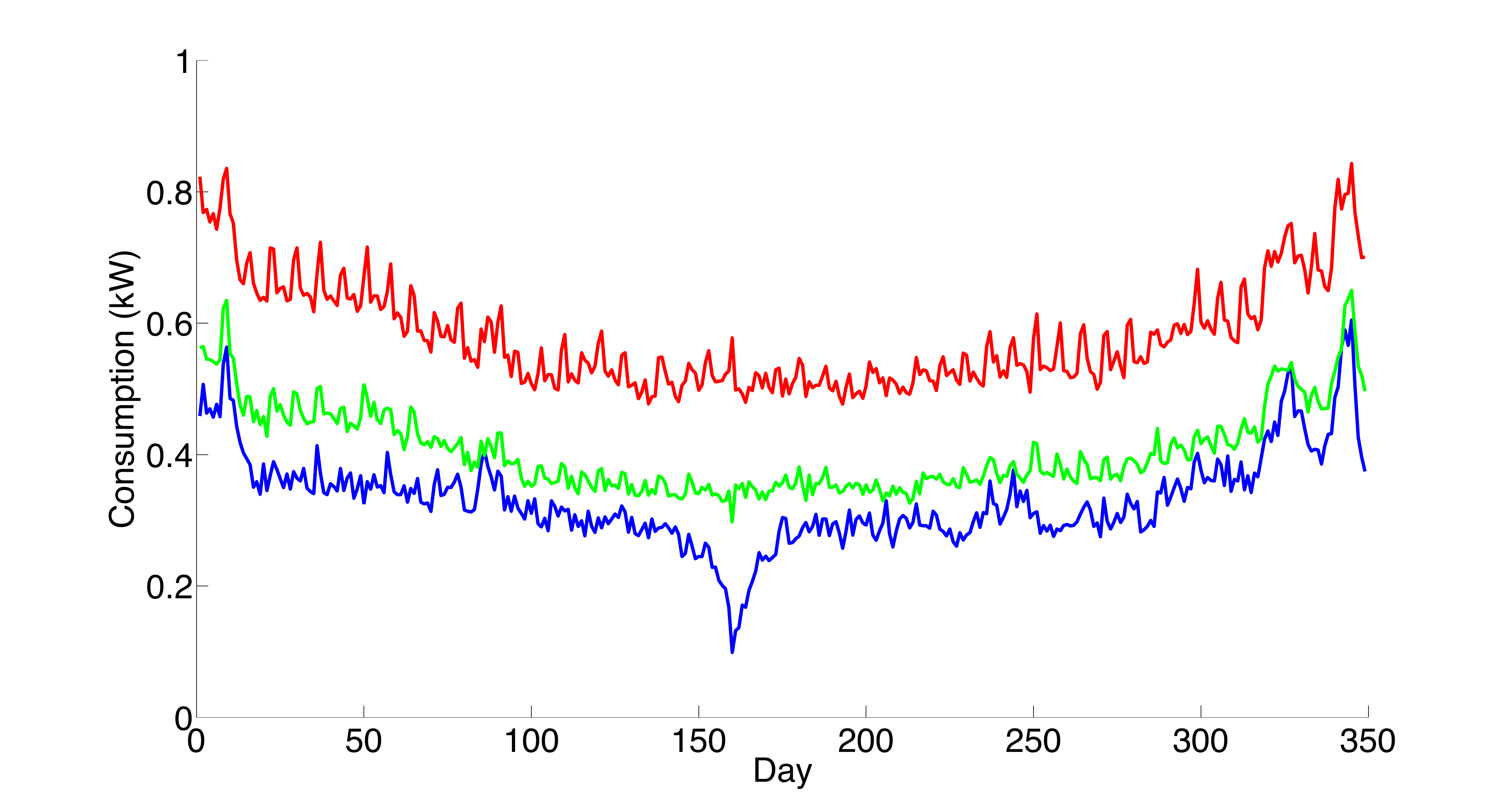}
 \caption{Daily mean consumptions of the cluster centres along the year for $3$ clusters, clusering done on weekend observation}
\end{figure}

 \section{Discussion and conclusion}
 
Massive information about individual (household, small and medium 
enterprise) consumption are now provided with new metering technologies 
and smart grids. Two major exploitations of individual data are load 
profiling and forecasting at different scales on the grid. Customer 
segmentation based on load classification is a natural approach for that 
and is a prolific way of research. We propose here a new methodology based 
on high-dimensional regression models. The novelty of our approach is that 
we focus on uncovering clusters corresponding to different regression 
models that could then be exploited for profiling as well as forecasting. 
We focus on profiling and show how, exploiting few temporal measurements 
of 500 residential customers consumption, we can derive informative 
clusters. We provide some encouraging elements about how to exploit these 
models and clusters for bottom up forecasting. 
 \bibliography{biblio}
 \bibliographystyle{plain}

\end{document}